\documentclass[iop]{emulateapj}

\usepackage[]{units}
\usepackage[usenames, dvipsnames]{color}
\usepackage{xspace}
\usepackage{bm}
\usepackage[normalem]{ulem}
\newcommand{\Lya}{Ly$\alpha\,\,$}
\newcommand{\lya}{Ly$\alpha$\xspace}

\newcommand{\unitcgslum} {erg\,s$^{-1}$\xspace}

\newcommand{\brithresh}{$\unit[4.45\times10^{-18}]{erg\,s^{-1}\,cm^{-2}\, arcsec^{-2}}$\xspace}
\newcommand{\nlaes}{183 }

\newcommand{\sodenspeak}{2.7\xspace}
\newcommand{\sodenstofield}{4.1\xspace}
\newcommand{\sodenspeakwerrors}{$2.7\pm1.1$\xspace}

\newcommand{\vodens}{10.4\xspace}

\newcommand{\rhovv}{$(6\pm1.5)\times10^{-4\,}\unit[]{Mpc^{-3}}$\xspace}
\newcommand{\rhos}{$(7.4\pm0.54)\times10^{-2\,}$arcmin$^{-2}$ $\Delta z^{-1}$\xspace}
\newcommand{\rhosv}{$(7.4\pm1.9)\times10^{-2\,}$arcmin$^{-2}$ $\Delta z^{-1}$\xspace}
\newcommand{\rhosfield}{$(5.4\pm0.9)\times10^{-2\,}$arcmin$^{-2}$ $\Delta z^{-1}$\xspace}
\newcommand{\peakrhos}{$0.27\,$arcmin$^{-2\,}\Delta z^{-1}$\xspace}

\newcommand{\rasize}{69.5}
\newcommand{\decsize}{35.6}
\newcommand{\xsize}{112.9\xspace}
\newcommand{\ysize}{57.9\xspace}
\newcommand{\zsize}{46.4\xspace}
\newcommand{\volume}{$\unit[3.03\times 10^5]{Mpc^3}$\xspace}

\newcommand{\vect}[1]{\mathbf{#1}}
\newcommand{\boo}{Bo\"otes\xspace}

\newcommand\h      {$^{\rm h}$}
\newcommand\m      {$^{\rm m}$}
\newcommand\s      {$^{\rm s}$}
\newcommand{\nb}   {{\sl NB}\xspace}
\newcommand{\bw}   {{\sl B}$_{\rm W}$\xspace}

\slugcomment{Submitted to ApJ (May 02). Revised (\today)}
\shorttitle{Proto-cluster at \lowercase{$z$}=2.3}
\shortauthors{B{\u a}descu et al.}

\begin{document}

\title{Discovery of a Proto-cluster associated with a L\lowercase{y}$\alpha$ Blob Pair at \lowercase{$z$}=2.3}

\author{Toma B{\u a}descu\altaffilmark{1,*},
        Yujin Yang\altaffilmark{1,2},
        Frank Bertoldi\altaffilmark{1},
        Ann Zabludoff\altaffilmark{3},
        Alexander Karim\altaffilmark{1}, 
        Benjamin Magnelli\altaffilmark{1}
}

\altaffiltext{1}{Argelander Institut f\"ur Astronomie, Universit\"at Bonn, Auf dem H\"ugel 71, 53121 Bonn, Germany}
\altaffiltext{2}{Korea Astronomy and Space Science Institute, 776 Daedeokdae-ro, Yuseong-gu, Daejeon 34055, Korea}
\altaffiltext{3}{Steward Observatory, University of Arizona, 933 North Cherry Avenue, Tucson AZ 85721}
\altaffiltext{*}{Email: toma@astro.uni-bonn.de}

\begin{abstract}

Bright \Lya blobs (LABs) --- extended nebulae with sizes of
$\sim$100\,kpc and \Lya luminosities of $\sim$10$^{44\,}$\unitcgslum
--- often reside in overdensities of compact \Lya emitters (LAEs)
that may be galaxy protoclusters.  The number density, variance,
and internal kinematics of LABs suggest that they themselves trace
group-like halos. Here we test this hierarchical picture, presenting
deep, wide-field \Lya narrowband imaging of a 1$^\circ$ $\times$
0.5$^\circ$ region around a LAB pair at $z$ = 2.3 discovered
previously by a blind survey.  We find \nlaes \Lya emitters, including
the original LAB pair and three new LABs with \Lya luminosities of
(0.9--1.3)$\times$10$^{43\,}$\unitcgslum and isophotal areas of
16--24 arcsec$^2$.  Using the LAEs as tracers and a new kernel
density estimation method, we discover a large-scale overdensity
(\boo J1430+3522) with a surface density contrast of $\delta_{\Sigma}$
= \sodenspeak, a volume density contrast of $\delta$ $\sim$ \vodens,
and a projected diameter of $\approx$\,20 comoving Mpc.  Comparing
with cosmological simulations, we conclude that this LAE overdensity
will evolve into a present-day Coma-like cluster with $\log{(M/M_\odot)}$
$\sim$ $15.1\pm0.2$.  In this and three other wide-field LAE surveys
re-analyzed here, the extents and peak amplitudes of the largest
LAE overdensities are similar, not increasing with survey size,
implying that they were indeed the largest structures then and do
evolve into rich clusters today.  Intriguingly, LABs favor the
outskirts of the densest LAE concentrations, i.e., intermediate LAE
overdensities of $\delta_\Sigma = 1 - 2$.  We speculate that these
LABs mark infalling proto-groups being accreted by the more massive
protocluster.

\end{abstract}

\keywords{
galaxies: clusters: individual (\boo J1430+3522) ---
galaxies: formation ---
galaxies: high-redshift ---
intergalactic medium ---
large-scale structure of universe
}

\section{Introduction}

The study of galaxy clusters plays an important role in understanding
cosmological structure formation and the astrophysics of galaxy evolution.
Statistics of galaxy cluster size, mass, and redshift distribution
provide constraints for cosmological models, while the properties of the
galaxies and gas inside clusters give clues about galaxy evolution and the
star formation history of the Universe \citep{press1974, lanzetta1995,
lilly1996, madau1998, boylan2009}. Progenitors of galaxy clusters, the
so-called proto-clusters, start off as overdense regions and groups of
galaxies at high redshift, which over time coalesce into the larger
galaxy clusters we see today. While galaxy clusters at $z$ $<$ 1 are
routinely discovered by various techniques such as the red sequence of
galaxies \cite[e.g.,][]{gyee2000, gyee2005}, the X-ray emission from
hot intracluster gas \citep{rosati2002, mullis2005, stanford2006}, or
the Sunyaev Zel'dovich effect on CMB photons \citep{sz1972, bleem2015},
observing the early stages of cluster formation at higher redshifts has
been challenging.

Since protoclusters lack many of the observational properties of massive
virialized galaxy clusters of today, one of the best ways to find them is
to identify galaxy over-densities at high redshift \citep{overzier2016}.
Readily observable populations of galaxies include radio galaxies
\citep{venemans2002, venemans2007, hatch2011radgal, hayashi2012,
wylezalek2013, cooke2014}, submillimeter galaxies \citep{daddi2009,
capak2011, rigby2014, dannerbauer2014}, Hydrogen Alpha Emitters (HAEs)
\citep{hatch2011haes, hayashi2012}, or  Lyman Break Galaxies (LBGs) and
Lyman Alpha Emitters (LAEs) \cite[e.g.,][]{taniguchi2005, overzier2006,
overzier2008}. LAEs, which are compact galaxies that have strong emission
in the Lyman-$\alpha$ line, are relatively easy to observe over a wide
range of redshifts at $z$ $\sim$ 2--6 \citep[e.g.,][]{taniguchi2005,
gronwall07, nilsson09, Guaita10}. LAEs are mainly star-forming, low
mass objects, and some may be the progenitors of today's Milky Way type
galaxies \citep{Gawiser2007}.  With wide-field, deep narrowband surveys
centered on the \Lya line emission at a given redshift, one can use
LAEs to identify galaxy overdensities. Giant \lya emitting nebulae, also
known as \lya ``blobs'' \cite[LABs;][]{Francis1996, Ivison1998, S2000,
P2004, M2004} which emit \Lya radiation on large scales (50--100\,kpc)
and have high \Lya luminosities of 10$^{43-44\ }$\unitcgslum are also
apparent tracers of LAE overdensities \cite[e.g.,][]{M2004, M2005,
Saito2006, P2008, YY2009, YY2010}.

What powers the strong extended \lya emission in blobs is still poorly
understood.  Possible powering mechanisms include gravitational
cooling radiation \citep{fardal01, haiman00, YY2006, dijkstra09,
FG2010, goerdt10, Rosdahl2012}, the resonant scattering of \lya photons
produced by star formation \citep{Moller1998, Laursen2007, hayes11,
steidel11, Zheng2011, Cen2013}, and photo-ionizing radiation from
active galactic nuclei (AGN) \citep{Haiman&Rees01, Cantalupo2005,
geach09, Kollmeier2010, Overzier2013, YY2014a}.  Another potential source is
shock-heating from starburst-driven winds \citep{taniguchi00, MU2006},
although recent studies of the emission of non-resonant lines from eight
\Lya blobs excludes models that require fast galactic winds driven by
AGN or supernovae \citep{YY2011, YY2014a, YY2014b, Prescott2015a}.

Regardless of the energy sources of \lya blobs, the association of
blobs with compact LAE overdensities with sizes of $\sim$10--20\,Mpc
\citep{M2004, M2011, P2004, P2008, YY2010, P2012, SM2014}, suggests that
LABs are good potential markers of large protoclusters. Furthermore,
the number density and variance of \Lya blobs, as well as the
200--400 km$\,$s$^{-1}$ relative velocities of their embedded
galaxies, suggests that blobs themselves occupy individual group-like
halos of $\sim$10$^{13}$M$_\odot$ \citep{P2012, YY2010, YY2011,
Prescott2015b}. Thus, blobs may be sites of massive galaxy formation and
trace significant components of the build-up of protoclusters. However,
because most previous LAB studies have been carried out toward known
over-dense regions or proto-clusters, the observed relationship between
\lya blobs and LAE overdensities may be biased.  To probe the LAB --
overdensity connection one should investigate the area around known
\Lya blobs that were identified without prior knowledge of their
environments. For example, \cite{P2008} studied the environment of a
\Lya blob that was serendipitously discovered by its strong Spitzer
MIPS $\unit[24]{\mu m}$ flux \citep{D2005}, finding that this \lya blob
resides in an over-dense region of 20$\times$50\,Mpc$^2$.

In this work, we investigate the large scale environment of a \lya
blob pair at redshift $z$=2.3 that was discovered without prior
knowledge of the environment \citep{YY2009}.  The paper is organized
as follows. In Section \ref{sec:obs}, we present our observations and
data reduction. In Section \ref{sec:analysis}, we discuss our selection
of \lya emitters and blobs.  In Section \ref{sec:resultsdiscussion}, we
describe the discovery of an overdensity associated with the \lya blob
pair, compare its properties with those obtained from three previous
narrowband surveys of other LAE structures, discuss whether it will
evolve into a present-day galaxy cluster, and show that \lya blobs are
preferentially located in the outskirts of proto-clusters here and
in the other surveys. In Section \ref{sec:conlusions}, we summarize
the results.  Throughout the paper, we adopt the following cosmological
parameters: $H_0=\unit[70]{km\;s^{-1}\;Mpc^{-1}}$, $\Omega_{\rm M}=0.3$,
and $\Omega_\Lambda=0.7$. All distances presented are in the comoving
scale unless noted otherwise, and all magnitudes are in the AB system
\citep{Oke1974}.

\section{Observations and Data Reduction }
\label{sec:obs}

\cite{YY2009} conducted a wide-field narrow-band survey covering an area
of $\unit[4.82]{deg^2}$ of the \boo\ NDWFS, targeting \Lya emission at
z=2.3, and obtained an unbiased sample of the largest and brightest \Lya
blobs at that redshift. The redshift was chosen to facilitate future
observations of the extended \Lya gas via the optically thin H$\alpha$
$\unit[6563]{\AA}$ line, which is redshifted into a relatively sky
line free part of the infrared spectrum. \cite{YY2009} discovered four
\Lya blobs with luminosities of $1.6-5.3\times10^{43}$\unitcgslum\ and
isophotal areas $28-57$ arcsec$^2$. Two of the four blobs form a pair,
with a separation of only 70" (550 kpc at z=2.3), which makes them
ideal targets for our deeper follow-up \Lya survey to map the spatial
distribution of LAEs and LABs described here.

\begin{figure}[t]
\centering
\includegraphics[width=0.55\textwidth]{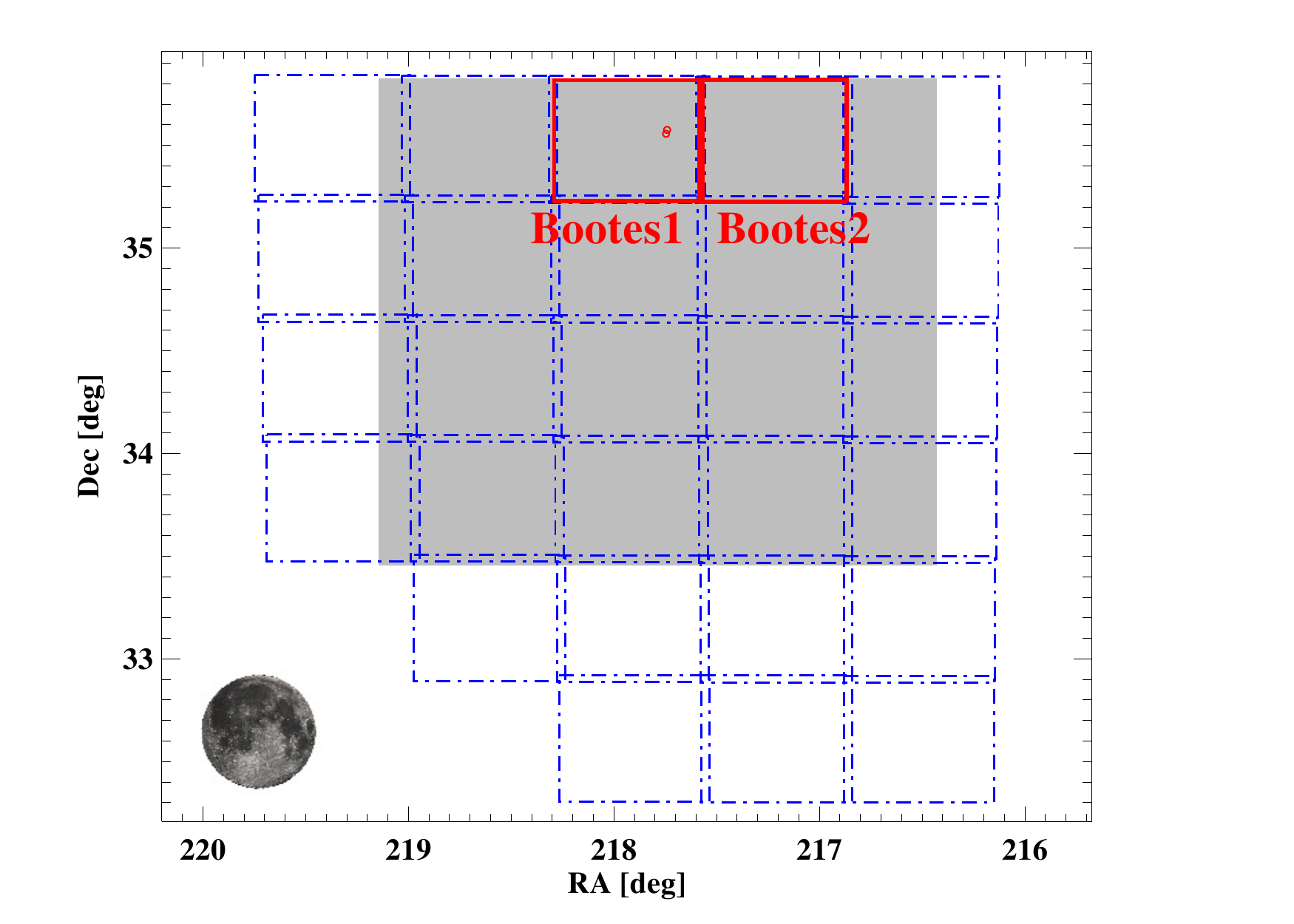}
\caption{The extent of the NOAO Deep Wide-Field Survey and the
previous \lya narrowband imaging survey \citep{YY2009} is shown in
blue rectangles and the shaded region in gray, respectively. The
two fields that we targeted in this work are indicated in red. The
\Lya blob pair discovered by \cite{YY2009} is marked with red
circles.  The total field of view of our new imaging has dimensions
of \rasize\arcmin$\,\times\,$\decsize\arcmin\ or \xsize Mpc $\times$
\ysize Mpc with a line-of-sight depth of $\approx$ \zsize Mpc. The Moon
is shown for scale.
}
\label{fig:ndwf3}
\end{figure}

We obtain narrowband images covering a total area of
$\sim$1\degr$\times$0.5\degr\ around the two known \Lya blobs
\citep{YY2009} using the Mosaic1.1 camera on the Kitt Peak National
Observatory (KPNO) Mayall 4m telescope.  In Figure \ref{fig:ndwf3}
we show the areas covered by the National Optical Astronomical
Observatory (NOAO) Deep Wide-Field Survey (NDWFS) and the locations
of our two pointings (hereafter Bo\"otes1 and Bo\"otes2) centered
on 14\h31\m42\fs22, +35\degr31\arcmin19\farcs9 and 14\h28\m54\fs08,
+35\degr31\arcmin19\farcs9. Observations were carried out on 2011 April
29 and 30, with exposure times of 7.3 and 6.0 hours, respectively. During
the two observing nights, the average seeing was $\approx$1\farcs1.

We observe with the custom narrowband filter used in the discovery of
the known \Lya blob pair \citep{YY2009}.  The filter has a central
wavelength of $\lambda_c=\unit[4030]{\mbox{\AA}}$ and a bandwidth
of $\Delta\lambda_{\rm FWHM}=\unit[47]{\mbox{\AA}}$, corresponding
to the \Lya emission at $z$ = 2.3 and a line-of-sight depth of
$\approx\unit[\zsize]{Mpc}$ ($\Delta$z=0.0037). Apart from the narrowband
(\nb) images, we also use NDWFS broadband \bw, $R$, and $I$ band images
for continuum estimation.

We reduce the data using the {\tt{MSCRED}} package in {\tt{IRAF}}
\citep{IRAF86}. We correct the images for cross-talk and bias, then
apply the flat-field correction, using both dome and sky-flats. Bad
pixels and satellite trails are masked, and cosmic rays are removed
using {\tt{LA-COSMIC}} software \citep{LACOSMIC}. We flux-calibrate by
observing 3--4 spectrophotometric standard stars per night, with typical
uncertainties in flux calibration of $\sim$0.02--0.04 magnitudes. The
astrometry of our images is improved with the {\tt{msccmatch}} tasks in
{\tt{IRAF}} using the USNO-B1.0 \citep{usnob1} catalog.  After matching
the image scales, we stack them using the {\tt{mscstack}} task. The total
field of view has dimensions of \rasize\arcmin$\,\times\,$\decsize\arcmin\
or \xsize Mpc $\times$ \ysize Mpc, with a total survey volume of \volume.

\section{Analysis}
\label{sec:analysis}
\subsection{Selection of \Lya Emitters}
\label{subsec:sellaes}

We run Source Extractor {\tt{SExtractor}} \citep{sextractor} on the \nb
image and select sources having at least four adjacent pixels above the
$1\sigma$ local background rms, identifying $\sim 45000$ sources. After
applying a $3\times3$ pixel ($0.768\times0.768$ arcsec) boxcar filter
to the \nb and \bw images, we extract the \nb and \bw magnitudes inside
circular 3\arcsec\ apertures centered on the selected sources. From these
we determine the \Lya line flux, equivalent width (EW), and underlying
continuum flux for each of our objects using the following relations:
\begin{eqnarray}
f_{cont}^{\lambda} &=& \frac{F_{Bw}-F_{N\!B}}{\Delta\lambda_{Bw}-\Delta\lambda_{N\!B}}\label{eq:fcont}\\
F_{line} &=& F_{N\!B}-\Delta\lambda_{N\!B}\cdot f_{cont}^{\lambda}\label{eq:fline}\\
B_W^{cont} &=& -2.5\log\left(f_{cont}^\lambda\frac{ \lambda_{N\!B}^2}{c}\right)-48.6,
\end{eqnarray}
where $f_{cont}^{\lambda}$ is the continuum flux density, $F_{line}$
is the \Lya line flux, $F_{N\!B}$ and $F_{Bw}$ are the fluxes in
the \nb and \bw bands respectively, and $\Delta\lambda_{N\!B}$,
$\Delta\lambda_{Bw}$ are the bandwidths of the two filters. $B_W^{cont}$
is the AB continuum magnitude, without the line contribution, and
$\lambda_{N\!B}$ = 4030\AA\ is the central wavelength of the \nb\ 
filter.

\begin{figure*}[ht]
\includegraphics[trim=1.4cm 2.2cm 0.4cm 2.5cm, clip=true,scale=0.3, width=0.48\textwidth]{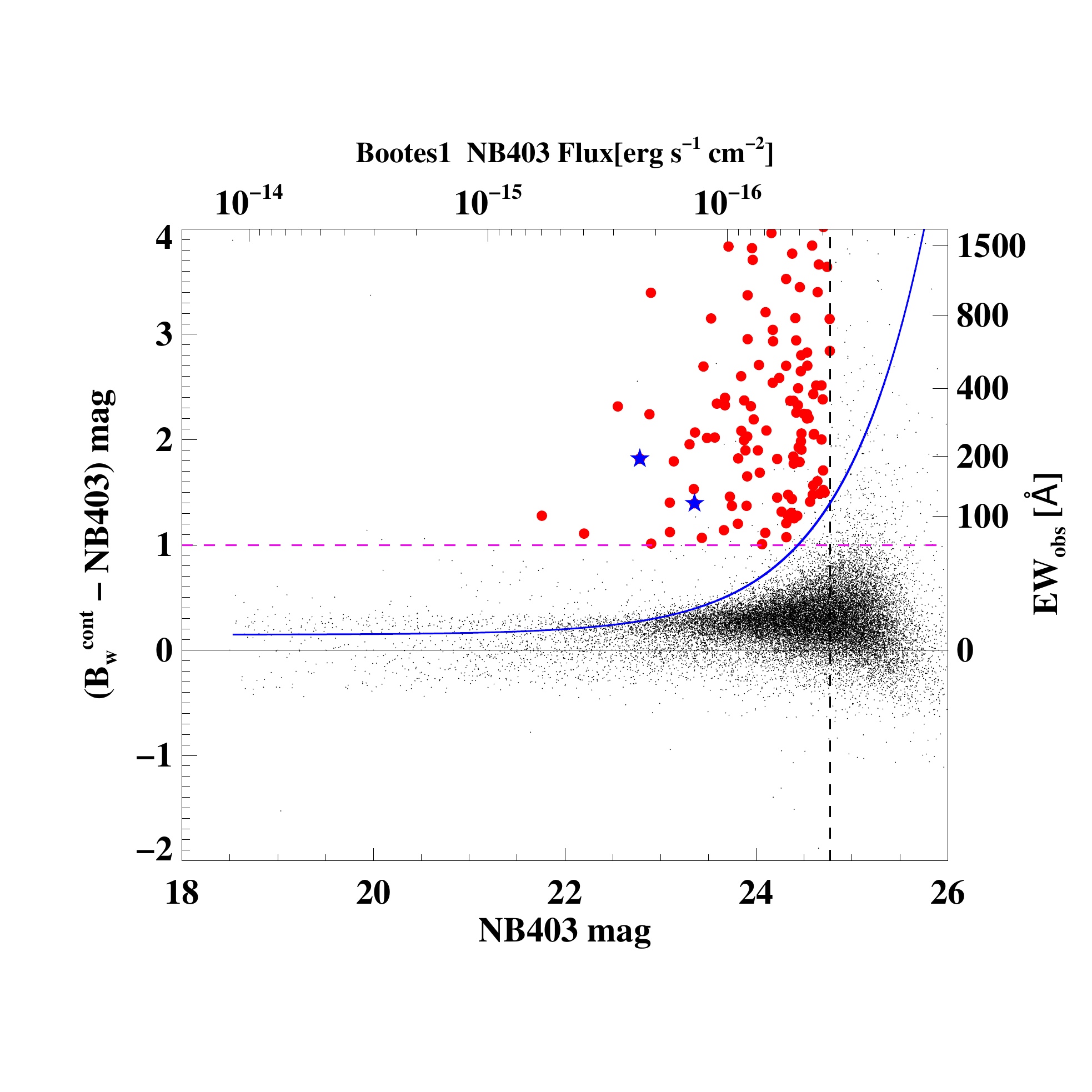}
\includegraphics[trim=1.4cm 2.2cm 0.4cm 2.5cm, clip=true,scale=0.3, width=0.48\textwidth]{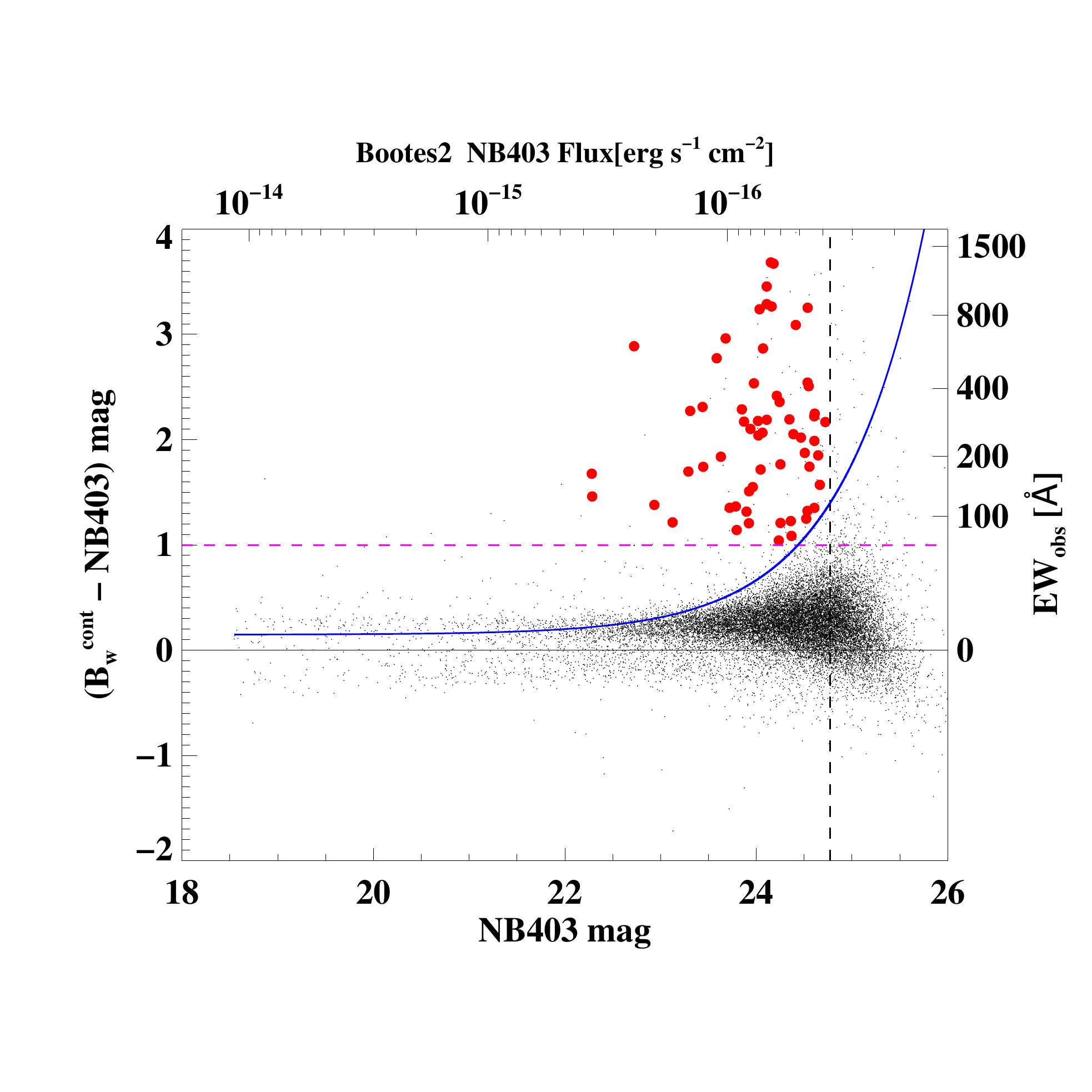}
\caption{
Color-magnitude plots of objects in the Bo\"otes1 (left) and Bo\"otes2
(right) fields (black dots). \Lya emitter candidates are marked
with red dots. The blue stars represent the two known blobs from
\cite{YY2009}. The horizontal dashed line marks the cut in EW$_{\rm
obs} > \unit[67]{\mbox{\AA}}$, while the vertical one represents the
cut in NB magnitude at 24.77. The blue curve represents the $5\sigma$
\nb magnitude error cut.
}
\label{fig:colmag}
\end{figure*}

To identify excess \Lya emission, we calculate the color index ($B^{\rm
cont}_{W}-N\!B$) of all our candidate sources. We create the \Lya
emitter sample by applying the following selection criteria to the
extracted objects:
\begin{itemize}
\item $B^{\rm cont}_{W}-N\!B \geqslant 1$, 
corresponding to ${\rm EW}_{\rm obs}\geqslant67{\rm \AA}$
\item $N\!B\leqslant24.77$ ($5.5\sigma$ detection threshold)
\item $B^{\rm cont}_{W}-N\!B \geqslant 5\sigma_{N\!B}$,
\end{itemize}
where $B^{\rm cont}_{W}$ is the continuum magnitude of
an object, without the \Lya line emission. The $5.5\sigma$
narrowband detection threshold corresponds to a \Lya luminosity of
$1.6\times10^{42\,}\unit[]{erg\,s^{-1}}$, which is $\approx$3 times
deeper than the original wide field survey \citep{YY2009}.

In Figure \ref{fig:colmag}, we show the \nb magnitude versus color
index and equivalent width for the \boo1 and \boo2 fields. The dashed
vertical and horizontal lines correspond to our selection criteria in
\nb magnitude and color, respectively. After applying these cuts, we are
left with a sample of 354 objects. The blue solid lines correspond to
the cut imposed requirement that the color index should be larger than
5 times the error in the \nb magnitude, which eliminates 77 objects
from our sample.  Removing objects that are close to bright stars or
less than 50 pixels away from the image edges further reduces the size
of the sample to 223 objects. Finally, we inspect the sample visually,
eliminating obvious false detections, like bright nearby galaxies or
image artifacts, producing a final sample of \nlaes objects.
We consider sample contamination from [\ion{O}{2}]\,$\lambda$3727 emission
in galaxies at $z\approx0.08$. The rest-frame EW of [\ion{O}{2}] emitters
at $z$ = 0.1 -- 0.2 is $<$ 50\AA\ \citep{HOGG98,Chiardullo2013}, below
our EW cut.  Given that [\ion{O}{2}] EWs e-fold with a scale length of
6\AA--14\AA\ \citep{Chiardullo2013}, we estimate that the probability
of finding [\ion{O}{2}] interlopers with EW$_{\rm obs}$ $>$ 67\AA\
is less than 1\%.
We list the properties of the \nlaes \Lya emitters in Table
\ref{tab:LAEsample}.

We test how our selection criteria might influence the size and spatial
distribution of our LAE sample. We create 81 different \Lya emitter
samples by varying the selection criteria around our original values.
We vary the color index cuts, from 0.8 to 1.2 in nine steps of 0.05,
and the \nb magnitude cuts, from 24.69 to 24.85 in nine steps of 0.02
mag. Comparing all the resulting samples to the one we originally adopted
for this work, we find that the influence of using these different
selection criteria on the large scale distribution of objects is minimal
(see Section \ref{subsec:largesc}).

\begin{deluxetable}{ccccc}
\tablecaption{Catalog of \Lya Emitter Candidates}
\tablehead{
\colhead{ID}&
\colhead{R.A. (J2000)}&
\colhead{Decl. (J2000)}&
\colhead{${\log(L_{\rm Ly\alpha})}$}&
\colhead{EW ($\unit[]{\AA}$)}
}
\startdata
1 &  14:32:36.39 & +35:23:34.7 & 42.41$\pm$0.05&  83\\
2 &  14:32:13.86 & +35:14:29.3 & 42.52$\pm$0.04& 113\\
3 &  14:30:27.02 & +35:14:32.7 & 42.17$\pm$0.08& 314\\
4 &  14:32:08.58 & +35:14:37.6 & 41.63$\pm$0.25& 137\\
... & ...        & ...       & ...             
\enddata
\tablecomments{This table is published in its entirety in the electronic
edition of the Astrophysical Journal.  A portion is shown here for
guidance regarding its form and content.}
\label{tab:LAEsample}
\end{deluxetable}

\begin{figure*}[t]
\centering
\includegraphics[width=0.47\textwidth ,clip=true,trim=1cm 0cm 0cm 0cm]{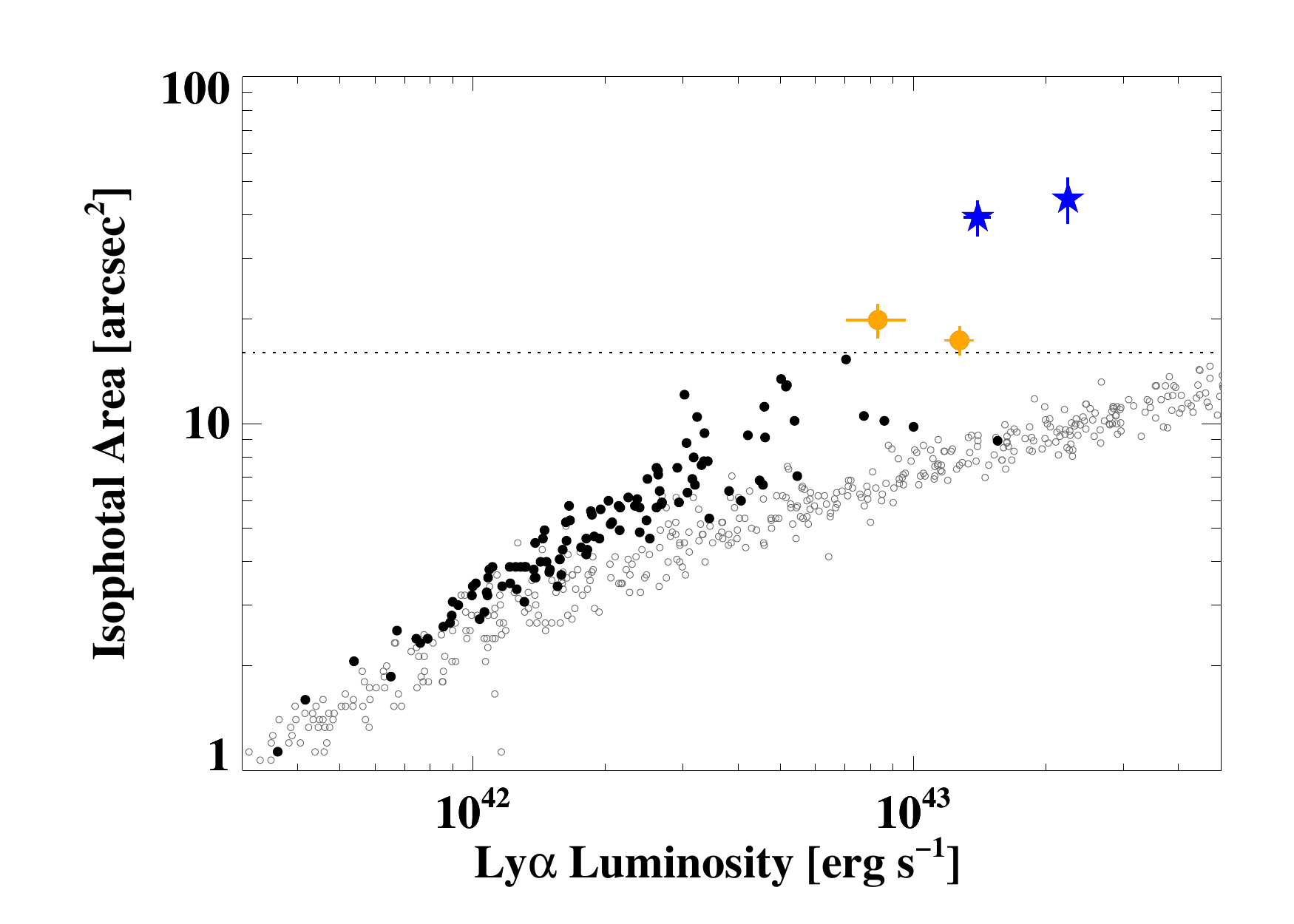}
\includegraphics[width=0.47\textwidth ,clip=true,trim=1cm 0cm 0cm 0cm]{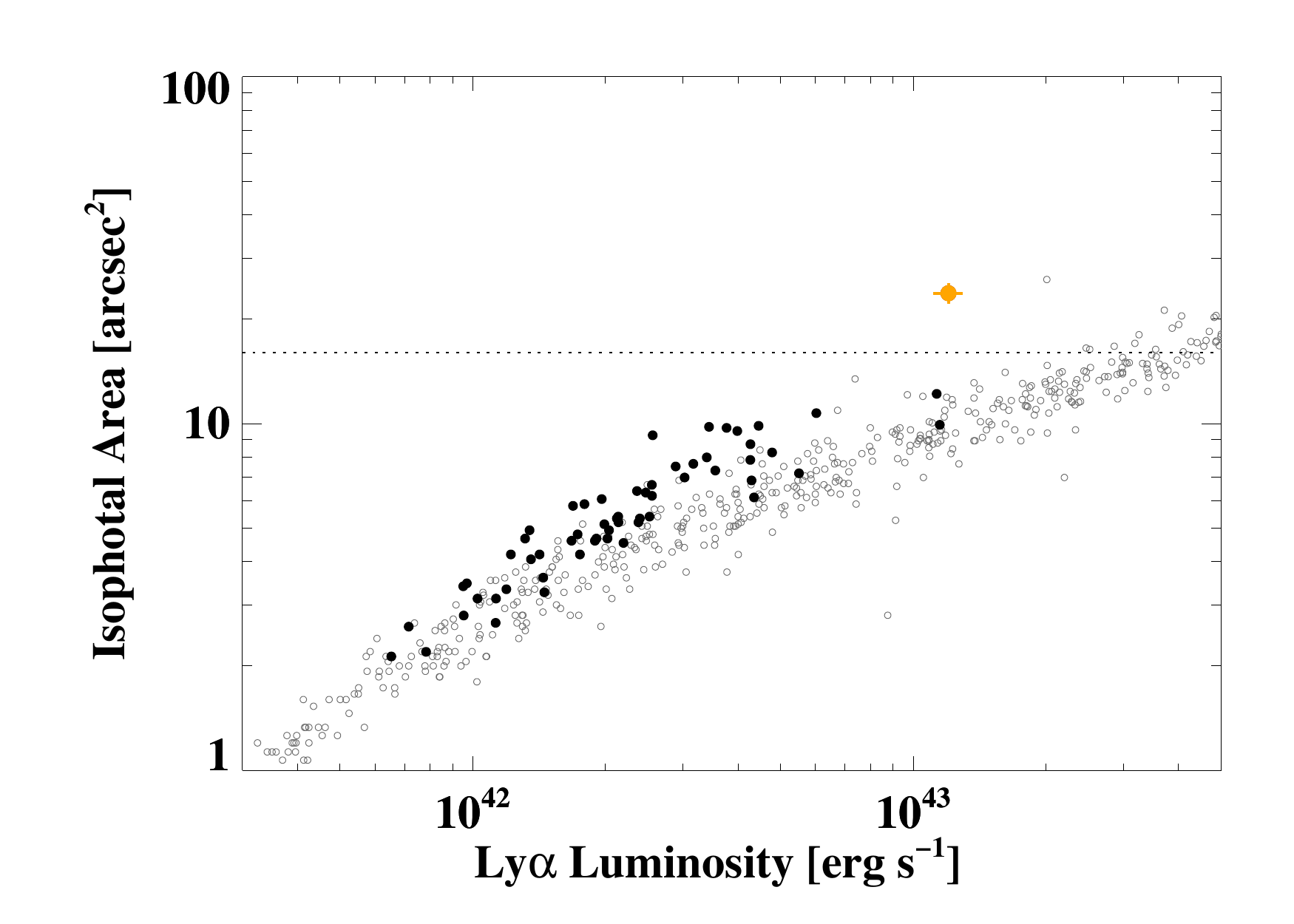}
\caption{
\Lya luminosity vs.\ the isophotal area of \Lya emitters, including
\Lya blobs, for the Bo\"otes1 field (left) and the Bo\"otes2 field
(right). \Lya emitters, the three new \Lya blob candidates, and the
two known blobs \citep{YY2009} are shown as black dots, filled orange
circles, and blue stars, respectively.  The dotted horizontal line marks
the selection criteria for our \Lya blobs: an isophotal area greater
than $\unit[16]{arcsec^2}$ above the \brithresh brightness limit. The
gray circles represent simulated point sources in our fields.
}
\label{fig:lumarea}
\end{figure*}

\subsection{Selection of \Lya Blobs}
\label{subsec:sellab}

With deeper \nb imaging data than those in \cite{YY2009}, we search
for \lya blobs with intermediate luminosities and sizes that our
prior shallower survey might have missed.  Using Eqs.~(\ref{eq:fcont})
and (\ref{eq:fline}), we calculate the \Lya line flux for each pixel.
The 1$\sigma$ surface brightness limit of the resulting \lya line image
is $\sim 2.1\times10^{-18}\unit[]{erg\,s^{-1}cm^{-2}arcsec^{-2}} $ per
1\,arcsec$^2$ aperture, which makes this survey 1.5--2.2 times deeper
than the original wide field survey that led to our discovery of the
LAB pair \citep{YY2009}. We run {\tt SExtractor} on the line image,
selecting sources with at least 16 adjacent pixels above the $5\sigma$
surface brightness limit. Then we cross-match this catalog with our
emitter sample above to make sure that the extracted \lya blob candidates
do have a \lya line excess.  We select \lya blob candidates by requiring
that their isophotal area above the surface \Lya brightness threshold of
\brithresh is larger than $\unit[16]{arcsec^2}$. We initially find seven
objects matching this criterion, including the two known blobs. In order
to estimate possible sample contamination, we place artificial point
sources having $L$(\lya) = $\unit[10^{41-44}]{erg\,s^{-1}}$ in our \Lya
images and extract them using the same procedures as for the LABs.

Because the noise and background level of the image can vary across the
field, we also  test how reliably we can recover extended \Lya emission
for the LAB candidates. We cut out 101$\times$101 pixels regions around
the candidates from the \Lya line image, centered on the candidates,
and place them in 4000 empty sky regions in the \boo 1 and 2 fields. We
then run the source extraction procedure using the same settings as
for the real data. The measured size and luminosities of the \lya blob
candidates will vary depending on the position in the field. The variance
of the source properties recovered this way gives us the uncertainties
on the luminosities and sizes of the candidates introduced by placing
the objects in different parts of the field. The recovery fraction is
defined as the fraction of times the \lya blob candidate is recovered
with a size above $\unit[16]{arcsec^2}$.  Out of the seven initial
candidates, five candidates --- including the already known blob pair
--- have recovery fractions higher than 90\%. We consider these to be
our LAB sample.  The 90\% recovery threshold was chosen because 
the rest of the recovered blobs have much lower recovery fractions:
two blobs with 75\% and the rest well below 50\% recovery fraction.
In Figure \ref{fig:lumarea}, we show the isophotal area of the \lya blob
candidates against their \Lya luminosity, as well as the relations for
the simulated point sources. The \lya blob candidates are located at
higher isophotal areas for a given luminosity, clearly separated from
the locus of point sources.

\begin{deluxetable*}{cccccc}
\tablecolumns{6} 
\tablecaption{Properties of \Lya Blobs}
\tablehead{ 
\colhead{Object} &
\colhead{R.A.} & 
\colhead{Decl.} & 
\colhead{$L$(Ly$\alpha$)} &
\colhead{Size} & 
\colhead{Recovery} \\
\colhead{}&
\colhead{(J2000)} & 
\colhead{(J2000)} & 
\colhead{($\unit[10^{43}]{erg \,s^{-1}}$)} &
\colhead{(arcsec$^2$)} & 
\colhead{fraction}
}
\startdata
\boo-LAB1 & 14 30 59.0 & +35 33 24 & $2.70\pm0.10$    & $43\pm4.8$   & --      \\
\boo-LAB2 & 14 30 57.8 & +35 34 31 & $1.61\pm0.08$    & $29\pm6.9$   & --      \\[0.2ex]
\hline\\[-1.5ex]
\boo-LAB5 & 14 32 17.7 & +35 47 53 & 1.360$\pm$0.098  & 15.5$\pm$1.7 & 97.5\%  \\ 
\boo-LAB6 & 14 30 50.2 & +35 41 03 & 0.931$\pm$0.129  & 16.1$\pm$2.3 & 96.2\%  \\ 
\boo-LAB7 & 14 30 13.0 & +35 37 45 &  1.261$\pm$0.094 & 15.6$\pm$1.6 & 93.9\%   
\enddata
\tablecomments{
We adopt a naming convention such that the four \lya blobs in the original
wide and shallow survey \citep{YY2009} are \boo-LAB1 to \boo-LAB4 and
that the three new \lya blobs found in this study are \boo-LAB5 to
\boo-LAB7. Note that \boo-LAB3 and \boo-LAB4 are not included in the
table because they were not covered by this new survey. Blobs are listed
in the order of recovery fraction.
}
\label{tab:blobs}
\end{deluxetable*}

\begin{figure}[h]
\centering
\textbf{Known \lya Blobs (Yang et al. 2009)}\par
\smallskip
\includegraphics[width=8.5cm,clip=true,trim=0cm 17cm 0cm 0cm]{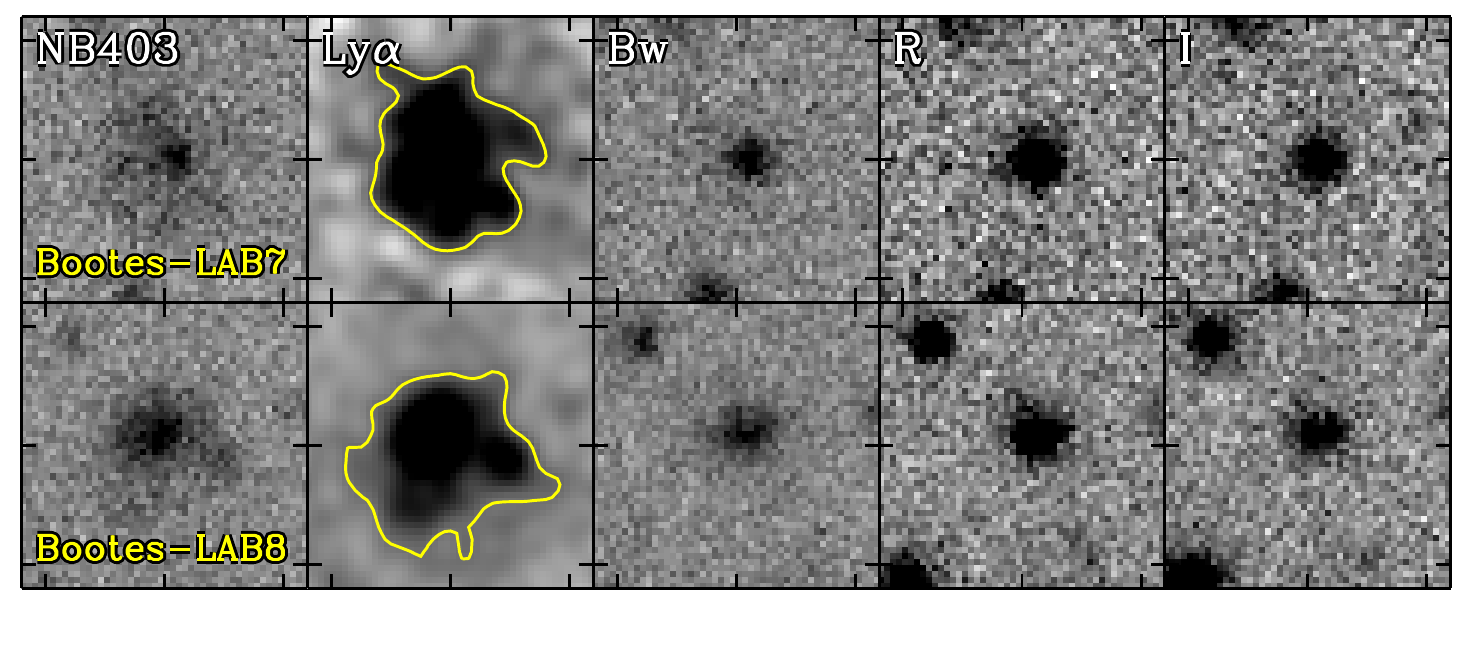}\\
\textbf{New Blobs in the \boo1 Field}\par
\smallskip
\includegraphics[width=8.5cm,clip=true,trim=0cm 17cm 0cm 0cm]{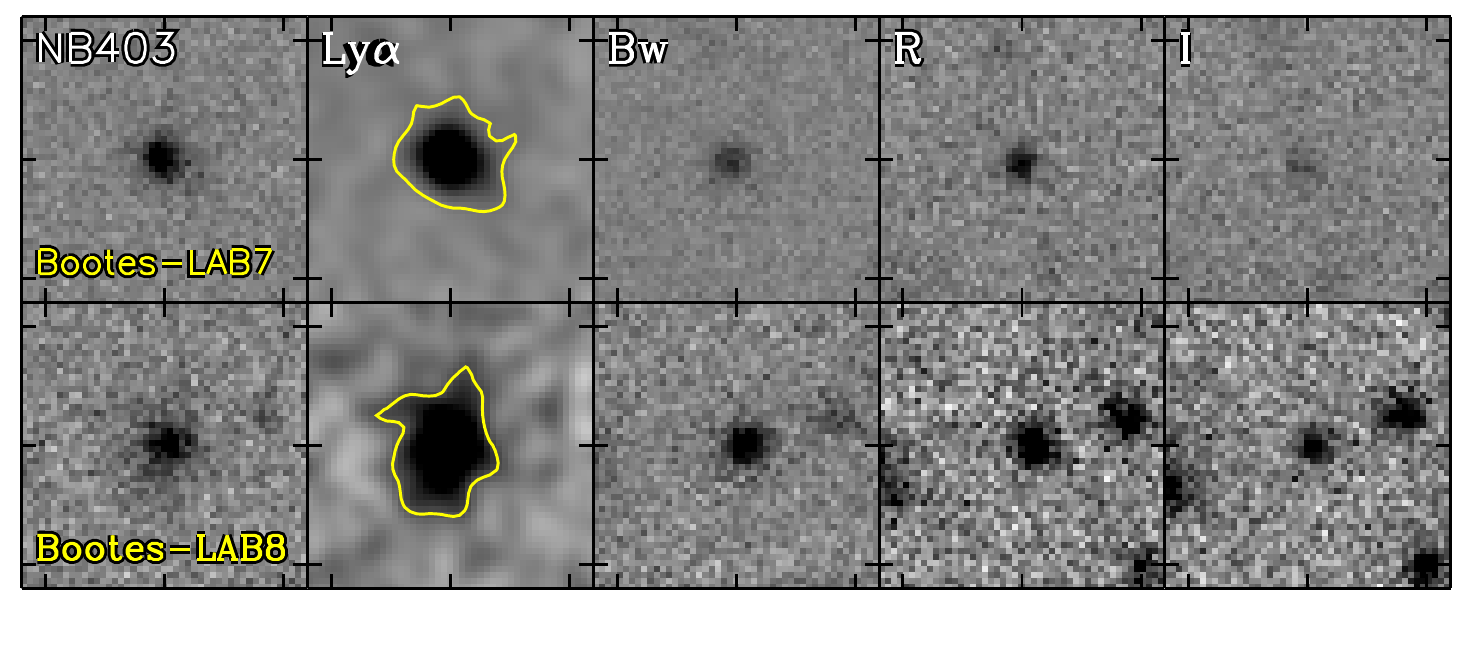}\\
\textbf{A New Blob in the \boo2 Field}\par
\smallskip
\includegraphics[width=8.5cm,clip=true,trim=0cm 20cm 0cm 0cm]{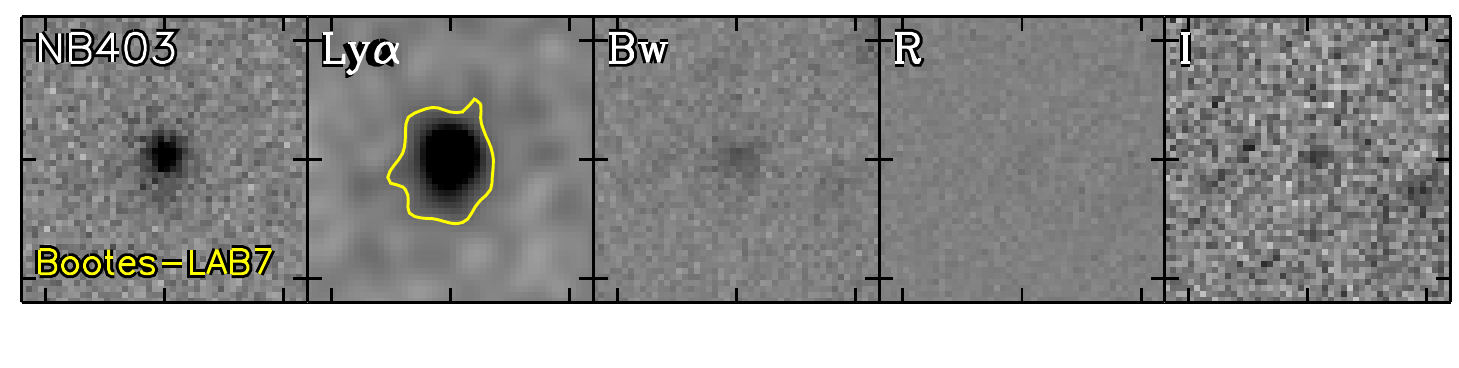}
\caption{
Images of the three new \Lya blob candidates (bottom three rows) and
of the two known blobs from \citet{YY2009} (top two rows). From left
to right: \nb, continuum subtracted \Lya image, $Bw$, $R$, and $I$
bands respectively.  The contours represent the surface brightness
of \brithresh. The isophotal areas of the new intermediate blobs are
$\approx$16 arcsec$^2$ with luminosities of $(0.9-1.4)\times10^{43}$
\unitcgslum. The distance between the tick-marks is $\unit[5]{arcsec}$.
}
\label{fig:7blobs}
\end{figure}

In Figure \ref{fig:7blobs}, we show all our \Lya blob candidates,
including the two known \Lya blobs of \cite{YY2009}, in the \nb,
\Lya line, \bw, $R$, and $I$ bands, respectively.  The shapes of the
\lya blob candidates are irregular and their isophotal areas exceed
those of their continuum counterparts.  In Table \ref{tab:blobs}, we
list the properties of the three new \Lya blobs, including position,
luminosity, and size.  Their \lya luminosities lie in the range of
$\unit[(0.9-1.4)\times10^{43}]{erg\,s^{-1}}$.

\section{Results and Discussion}
\label{sec:resultsdiscussion}

\subsection{Discovery of a LAE-traced Proto-cluster Associated with LABs}
\label{subsec:largesc}

\begin{figure*}
\begin{center}
\includegraphics[height=6.5cm,trim=0.5cm 0.5cm 0cm 4cm,clip=true]{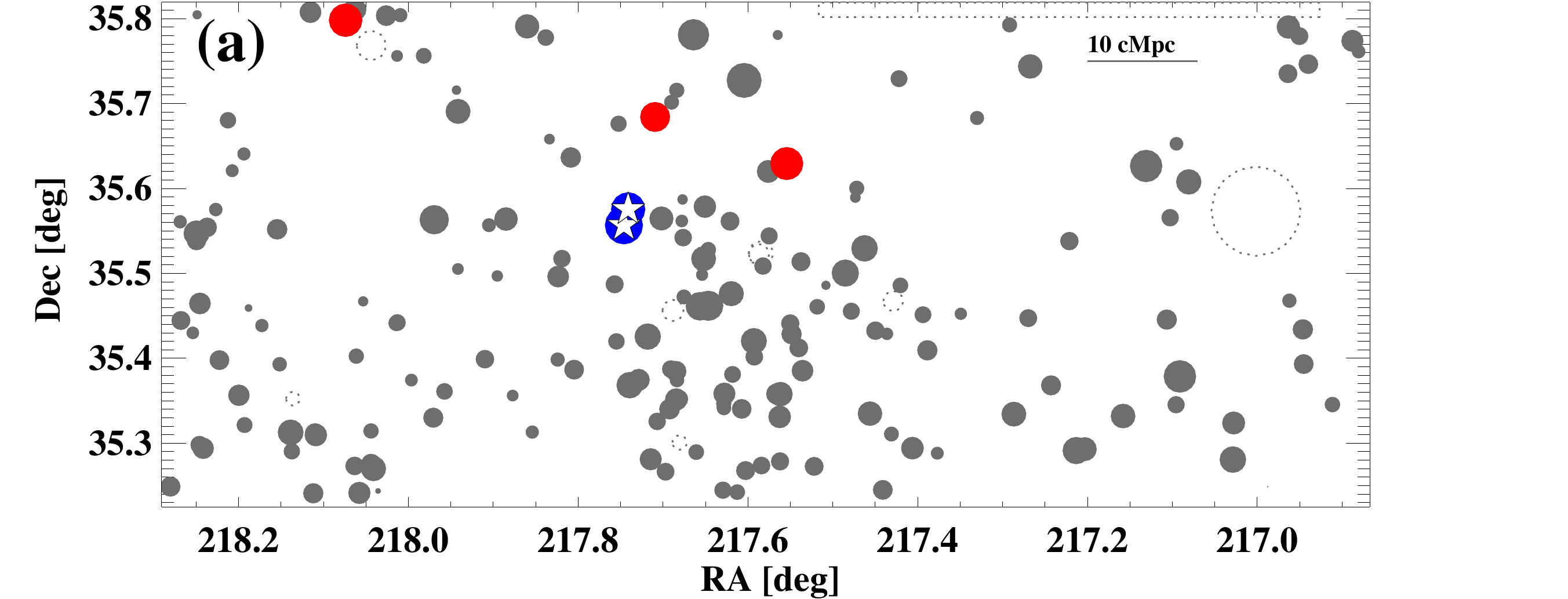}\\
\includegraphics[height=6.5cm,trim=0.5cm 0.5cm 0cm 4cm,clip=true]{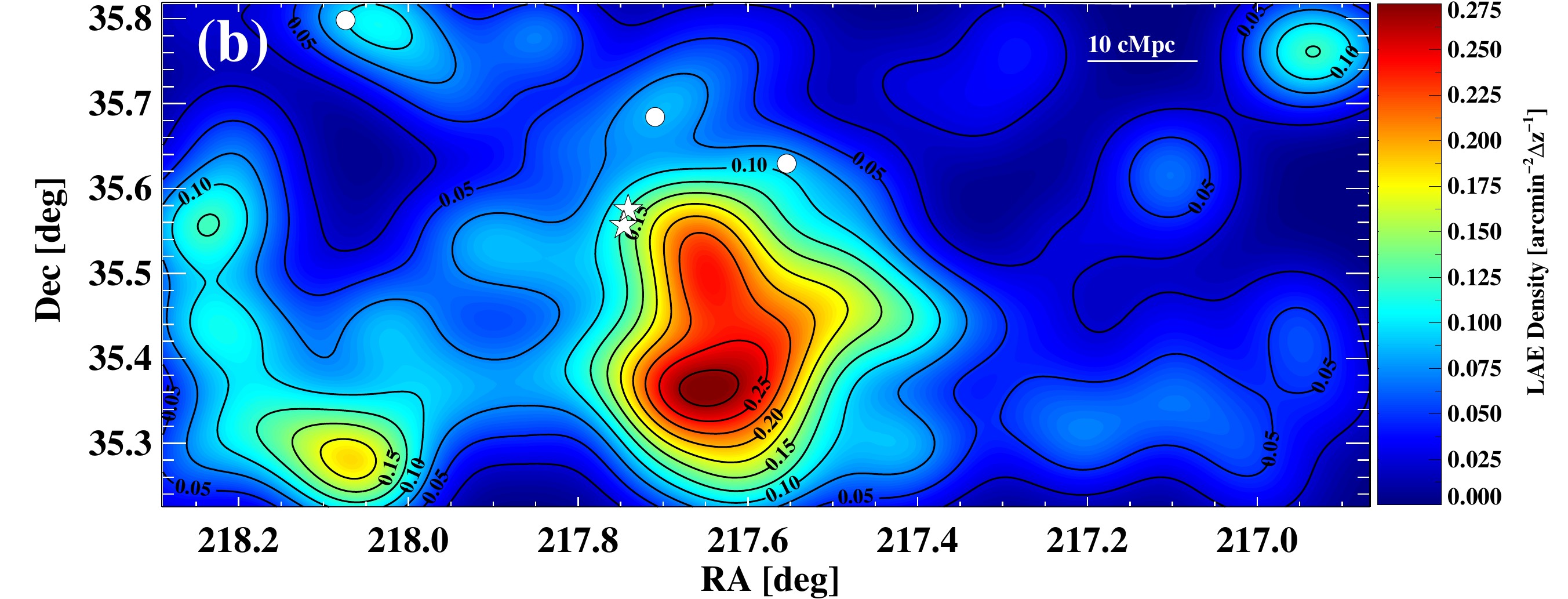}\\
\includegraphics[height=6.5cm,trim=0.5cm 0.5cm 0cm 4cm,clip=true]{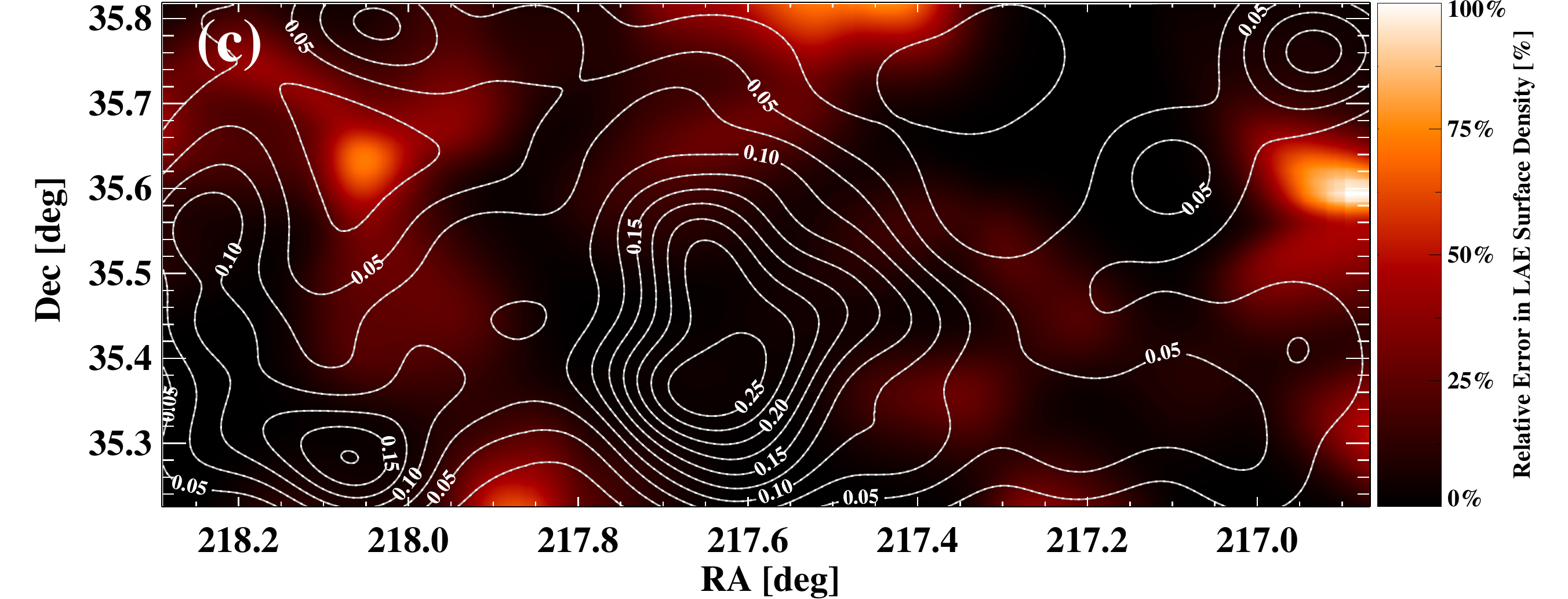}
\end{center}
\caption{
\textbf{(a)} Spatial distribution of \nlaes \Lya emitters and
blobs. Filled gray and red circles represent \Lya emitters and
new \Lya blob candidates, respectively. Star symbols are the
two previously known \lya blobs \citep{YY2009}. The radii of the circles are
proportional to the logarithm of the \Lya emitters' luminosities, in
the range of $\unit[10^{41.4-43.4}]{erg\,s^{-1}}$. The field of view
is 69.3\arcmin$\times$35.5\arcmin\ (138.5\,Mpc$\times$57.5\,Mpc). The
dotted lines enclose areas that have been excluded from our analysis
because of contamination from bright stars or galaxies.
\textbf{(b)} \Lya emitter surface density distribution obtained from the
KDE method explained in the Appendix. The contour labels show the surface
density of \Lya emitters in $\unit[]{arcmin^{-2}}$ per $\Delta z=0.037$
--- the value given by the narrowband filter width. The two known \Lya
blobs are marked with stars. The overdense region is clearly visible
towards the center of the image.
\textbf{(c)} Average and scatter of the \Lya emitter surface density
of the 81 surface density maps corresponding to different selection
methods.  The contour labels represent the average surface density while
the background image represents the scatter around this average map,
in percentages. The scatter is largest away from the overdense region,
increasing the confidence that the shape and size are not significantly
affected by varying the selection criteria.
\label{fig:radecmap}
\label{fig:contour}
\label{fig:contourerr}
}
\end{figure*}

Using our \nlaes \Lya emitter and blob sample, we investigate the large
scale environment around the  known \Lya blob pair \citep{YY2009}. In
Figure \ref{fig:radecmap}$a$, we show the spatial distribution of our
\Lya emitters --- which includes the new \Lya blobs --- across the
69.3\arcmin$\times$35.4\arcmin\ field. We mark the locations of our
\nlaes \Lya emitters, and indicate the areas that were excluded from
our analysis because of contamination from bright sources such as stars.

To estimate the smooth surface density distribution from the discrete
positions of the detected galaxies, one often convolves the position
map with a Gaussian kernel of width $\sigma$. The width of this kernel
affects the resulting surface density distribution, yet there is no single
way of selecting the smoothing method and size of a smoothing kernel. The
kernel size is often chosen to match the mode \citep{SM2014} or the median
\citep{M2011} of the distances between objects in a sample. \cite{M2005}
select a kernel width that matches the redshift dispersion introduced by
the peculiar velocity dispersion of their LAE sample, and \cite{YY2010}
use an adaptive kernel technique to smooth their LAE sample.  In this
paper, we choose a different approach, one meant to find the kernel size
generating the smoothed density field that has the highest probability
of representing our LAE sample. This technique is described in detail
in the Appendix, and we briefly explain it here.

Assuming our LAEs' positions are randomly drawn from an unknown underlying
density distribution $f$, we use kernel density estimation (KDE) to
find an estimate $\hat{f}$ for the density distribution function. Our
method involves convolving the discrete object map with Gaussian kernels,
generating smooth density maps. Each map is generated using a different
kernel width $\sigma$. We search for the $\sigma$ value that maximizes
the likelihood to observe our \Lya emitter sample, given the density
distribution estimate $\hat{f}$. We find this optimum value for the kernel
width to be $\sigma$ = 2.63\arcmin, which is used for the smoothed image
in Figure \ref{fig:contour}$b$.

The \Lya emitter density map in Figure \ref{fig:contour}$b$ reveals a
significant over-density near the field center (R.A. =14\h 30\m35.7\s,
decl.=+35\degr22\arcmin06\farcs2), with a projected radius of
$\sim$10\,Mpc. This over-dense region is in both Bo\"otes1 and 2 image
frames, and so is unlikely to be caused by different observing conditions
of the two fields or sample selection criteria.

To test if the surface density maps change due to the different selection
criteria, we create 81 surface density maps, each one corresponding to a
different cut in color index and \nb magnitude as described in Section
\ref{subsec:sellaes}. Figure \ref{fig:contourerr}$c$ shows the mean
and variance of the \Lya emitter surface density of these 81 maps. The
average density map shows an overdensity that is very similar in size
and position to the one we obtained using our selection criteria. The
variance is largest away from the overdense region, indicating that the
overall number density and density contrast of the overdense region is
not strongly dependent on the LAE selection criteria.

\begin{figure}
\centering
\includegraphics[width=0.48\textwidth]{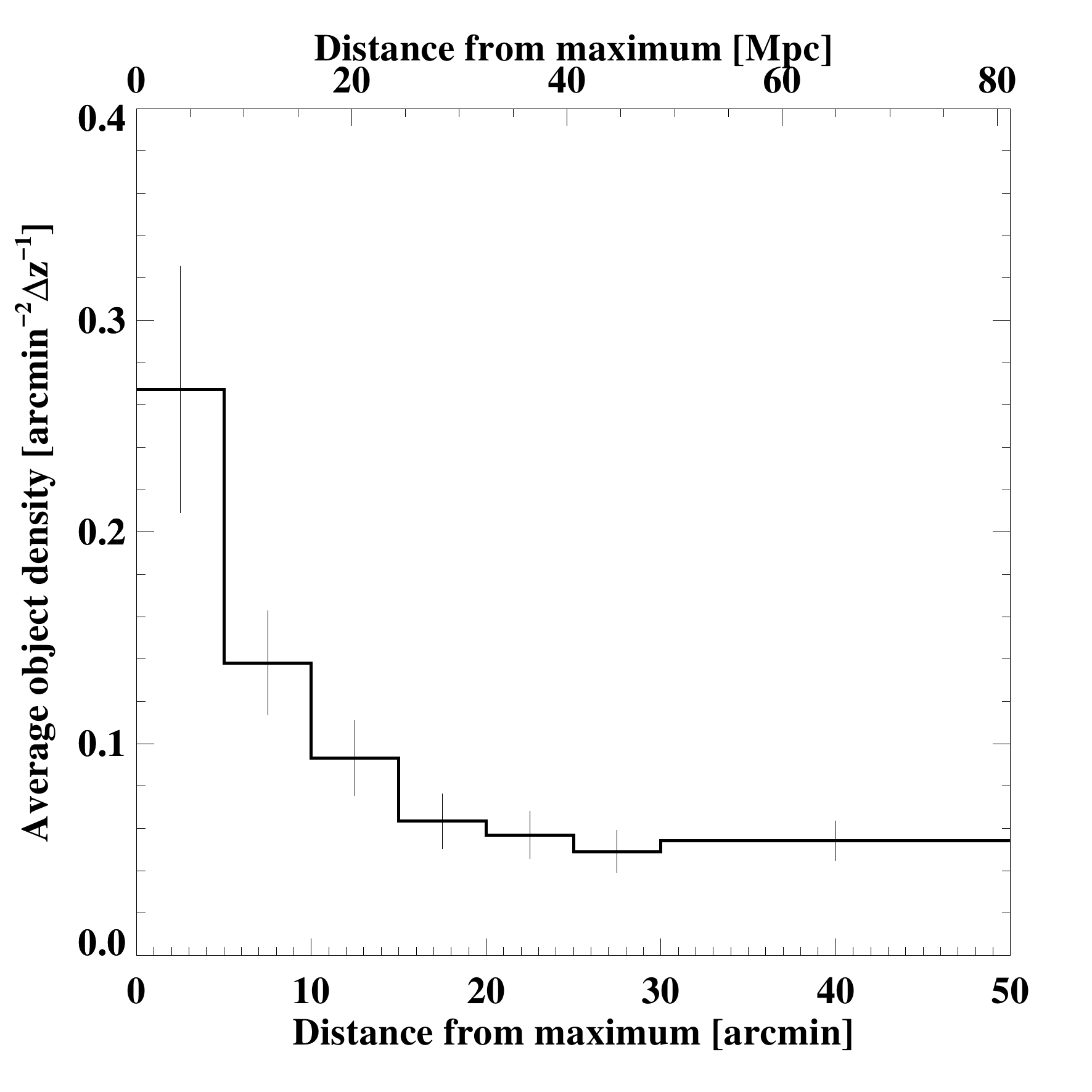}
\caption{
The radial profile of the surface density peak, as a function of distance
from the center of the over-dense region, using 5\arcmin\ bins for the
denser regions, and one 20\arcmin\ bin at the edge of the field. The peak
surface density is \peakrhos, decreasing to the field value of \rhosfield
at $r$ $>$ 30\,Mpc (20\arcmin) from the peak. The average surface density
over our entire survey is \rhos. The overdense region
has a radius of 10\,Mpc, a surface density contrast of $\delta_{\Sigma}$
= \sodenspeak, and a volume density contrast value of $\delta$ = \vodens.
\label{fig:radens}}
\end{figure}

To illustrate the size of the over-dense region, we show the radial
distribution of \Lya emitters in Figure \ref{fig:radens}.  The \Lya
emitter surface density peaks at $\Sigma_{\rm overdense}$ $\sim$ \peakrhos
inside a $\sim$8\,Mpc (5\arcmin) radius centered on the over-dense
region, decreasing to $\Sigma_{\rm field}$ = \rhosfield at radii larger
than 25\,Mpc, with an average value of $\bar{\Sigma}$ = \rhos over the
entire survey. The scale of this structure clearly demonstrates that one
needs a very wide field survey over $\sim$100\,Mpc to reliably measure
the overdensity relative to the background field region.

All uncertainties for the density measurements so far were
calculated assuming only Poissonian noise with a sample variance
$\sigma_{N}=\sqrt{N}$, where $N$ is the number of galaxies.  Cosmic
variance (CV) due to galaxy clustering can exceed sample variance and is
dependent on the survey geometry.  Although our survey volume is quite
large and the CV might be not significant, we also provide the density
measurements with uncertainties arising from cosmic variance.

We use the Cosmic Variance Calculator \citep{trenti08} to estimate
the CV for our survey volume. Given our survey configuration and a
sample completeness of 95\%, assuming a halo filling factor of 1,
we obtain a relative error due to cosmic variance of 25.7\% for our
survey geometry. The fractional error due to Poissonian shot noise
is 7.4\%.  Adding these errors in quadrature, the resulting relative
error is approximately 26.7\%.  Taking this error into consideration,
the average surface density over the whole field is \rhosv.

\subsection{Comparison with Previous Wide-Field LAE Surveys} 
\label{comparison}

We compare our LAE number densities with those of other surveys at
similar redshifts \citep{P2004,P2008,nilsson09,Guaita10,Mawatari12}.
Since each survey employs different selection criteria in EW and \nb
magnitude (\lya luminosity), as well as probing different redshift depths
due to different filter widths, we need to correct the reported LAE
surface density values in the literature. We scale the LAE surface
densities assuming \lya luminosity functions $\phi(L)$ at $z=2-3$ and
an exponential EW distribution ($e^{-w/w_0}$) with the scale length of
$w_0$. We calculate the following correction factors for each survey:
\begin{eqnarray}
C_{L}        &=& \frac{\int_{L_0}^{\infty}\phi(L')dL'}{\int_{L_i}^{\infty}\phi(L')dL'}\\
C_{\rm EW}   &=& \frac{\int_{{\rm EW}_0}^{\infty}\exp{(-w'/w_0)}dw'}{\int_{{\rm EW}_i}^{\infty}\exp{(-w'/w_0)}dw'}\\
C_{\Delta z} &=& {\Delta z_0}/{\Delta z_i},
\end{eqnarray}
where $C_{L}$, $C_{\rm EW}$, and $C_{\Delta z}$ are the correction
factors for \lya luminosity, equivalent width, and redshift
depth, respectively, $L_i$, ${\rm EW}_i$, and $\Delta z_i$ are
the luminosity limits, equivalent width cuts, and redshift depths
for different surveys, respectively; and $L_0$, ${\rm EW}_0$,
and $\Delta z_0$ are the values used in our survey. We adopt the
results from \cite{gronwall07} for the Schechter function assuming
no redshift evolution: $L^*=10^{42.66}\unit[]{erg\,s^{-1}}$,
$\Phi^*=1.28\times10^{-3}\unit[]{Mpc^{-3}}$, and $\alpha$ = $-1.36$,
and $w_0=74\unit[]{\AA}$. We summarize the results from the previous LAE
surveys, the adopted correction factors, and the LAE surface densities
corrected to our survey properties in Table \ref{tab:comparedens}.

\begin{figure}
\centering
\includegraphics[width=0.50\textwidth]{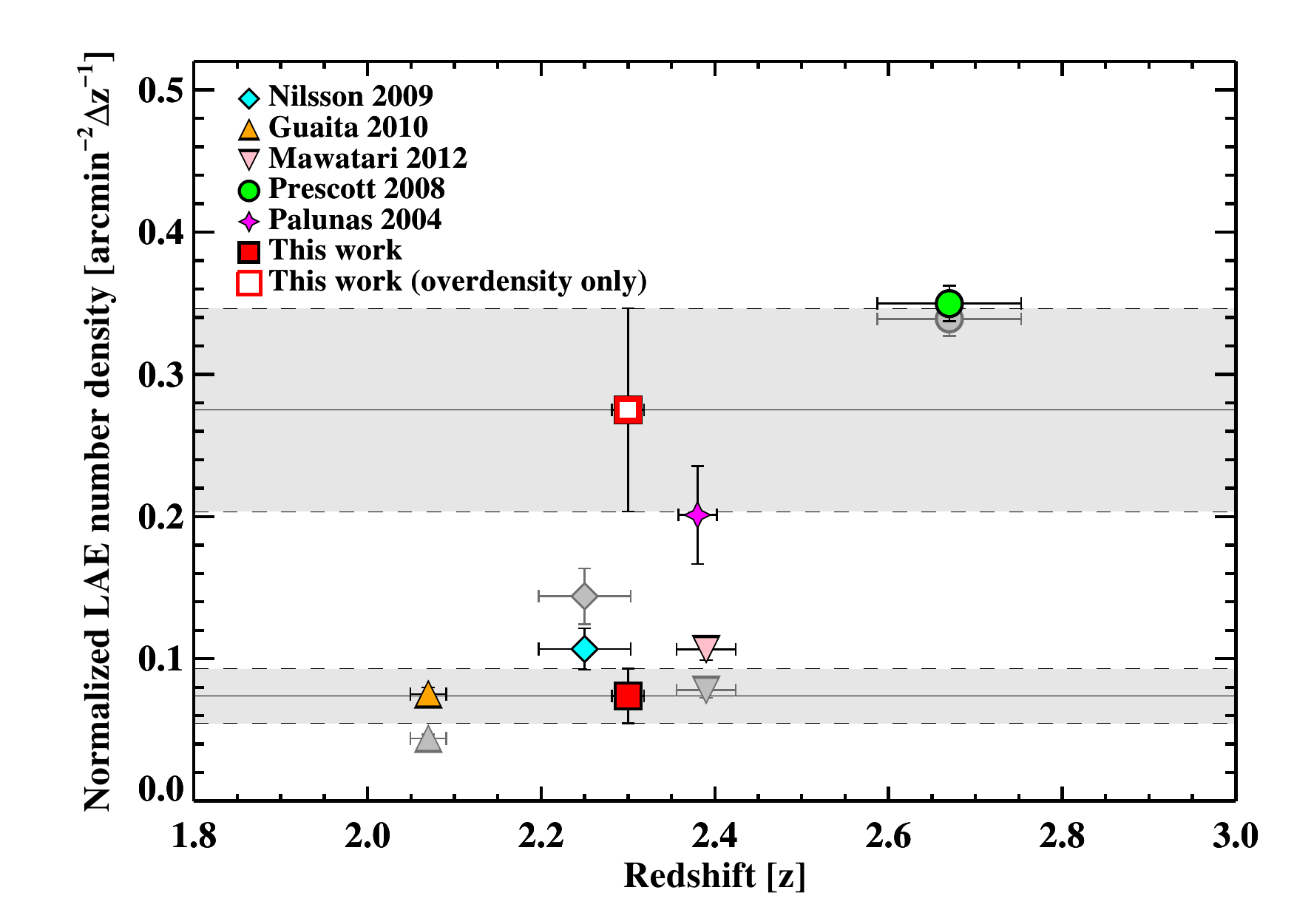} 
\caption{
Surface densities of LAEs, after correcting for the different selection
criteria of each survey, using the $z$=3.1 luminosity function
from \cite{gronwall07}. The gray symbols represent the same surface
densities corrected instead using the $z$=2.1 luminosity function of
\cite{Guaita10}. Since our survey extends well beyond the overdense
region, we consider the average surface density over the whole field
(filled square) and for the overdensity (open square) separately. The
gray bands show the 1$\sigma$ range about each value. Note that our
overdensity is consistent with that of known dense regions targeted by
\cite{P2004} and \cite{P2008}.
}
\label{fig:comparedens}
\end{figure}

In Figure \ref{fig:comparedens}, we show the surface density values
from other surveys, with redshifts close to $z$ = 2.3, and compare their
measurements with our peak and average surface densities.  Our average
surface densities agree with those of \cite{nilsson09}, \cite{Mawatari12},
and \cite{Guaita10}. The LAE density of our overdense region with a
radius of 10\,Mpc is in agreement with average density values from the two
surveys that targeted known dense regions, \cite{P2004} and \cite{P2008},
who targeted the J2143--4423 proto-cluster at $z$ = 2.38 and the LABd05
proto-cluster \citep{D2005}, respectively.

Given that the LF and EW distribution might evolve between $z$=2 and 3
\citep{Ciardullo2012}, we also test how the correction factors $C_L$,
$C_{EW}$, $C_{\Delta z}$ might be affected by the redshift evolution
of the luminosity function. We repeat the previous comparison
using the luminosity function of \cite{Guaita10} for $z$ = 2 with
$L^*=10^{42.33}$\unitcgslum, $\Phi^*=0.64\times10^{-3}$ Mpc$^{-3}$,
and $\alpha=-1.65$.  The resulting surface density values differ by
$\sim$30\% on average and by at most 70\% from the values in Table
\ref{tab:comparedens} (Figure \ref{fig:comparedens}).  Note that we do
not show the values for the shallowest \cite{P2004} survey because its
sources populate only the bright end of the luminosity function, which
introduces large errors when extrapolated to the faint end.

\subsection{Measurement of Surface and Volume Overdensity}

To gauge the significance of the discovered overdense structure, and to
compare its properties with the cosmological simulations and other known
protoclusters, we estimate the surface and volume \textit{over}-density
in this section.

The surface density contrast $\delta_\Sigma=(\Sigma_{\rm
overdense}-\bar{\Sigma})/\bar{\Sigma}$ is \sodenspeak inside a
8.1\,Mpc (5\arcmin) radius around the position of peak density. This
value increases to $\delta_\Sigma =(\Sigma_{\rm overdense}-\Sigma_{\rm
field})/\Sigma_{\rm field} =$ \sodenstofield if we compare our overdense
region to the {\it field} density ($\Sigma_{\rm field}$).
Throughout the paper and to be consistent with the definition of density
contrast used in the literature, we use the average density of the whole
survey (i.e., $\bar{\Sigma}$) when calculating overdensities. Calculating
contrast densities instead using the average field (i.e., $\Sigma_{\rm
field}$) value would increase the peak overdensity, while the standard
definition yields a more conservative result.

Assuming that the overdense region is a sphere with a radius of
10\,Mpc, we can estimate the volume density contrast as follows: We
find 35 LAEs inside a projected area with a 10\,Mpc radius centered
on R.A. = 14\h 30\m31.3\s, decl. = +35\degr25\arcmin01, while only
$\sim$9 LAEs are expected given the average volume density over the
survey.  Thus, we estimate that $\approx$26 more LAEs are located
within the assumed spherical overdensity having volume of $\approx$
4.18$\times$10$^3$\,Mpc$^3$. Our survey contains \nlaes objects
in a volume of $\unit[3.0\times10^5]{Mpc^3}$. The volume density
contrast $\delta = (\rho_{\rm overdense}-\bar{\rho})/\bar{\rho}$ is
then $\sim$\,10.4, where $\rho_{\rm overdense}$ is the density inside
the spherical region, and $\bar{\rho}$ is the average density over the
whole survey.

Several other surveys also find LAE overdensities at $z$ $\simeq$
2 -- 4. At $z$ = 2.16, a protocluster with an overdensity of
$\delta_{\Sigma}\sim 3$ is associated with the PKS 1138--262 radio galaxy
and its extended \Lya halo \citep{kurk2000b, venemans2007}.  Targeting
a known cluster J2143--4423 at $z$ = 2.38, \cite{P2004} find an LAE
overdensity of $\delta_{\Sigma}\simeq 2$. A similar surface overdensity of
$\delta_{\Sigma}\sim 2$ is seen by \cite{P2008} around a known LAB at $z$
= 2.7. An overdensity is found by \cite{SM2014} around the radio galaxy
TN J1338-1942 at redshift 3.1, with $\delta_{\Sigma}=2.8\pm0.5$.  At $z$
= 3.78, two or three over-densities with similar $\delta_{\Sigma}$ values
are found by \citet{Lee2014, Dey2016}, with $\delta_{\Sigma}=2.5-2.8$.
The largest overdensity by far lies in the SSA22 field \citep{S2000},
with a surface density contrast $\delta_{\Sigma}=5\pm2$ \citep{S2000,
M2004, M2005, yamada2012}.
More recently, \cite{Cai2016a} discover a massive overdensity at $z$ =
2.3, having a spectroscopically confirmed volume density contrast
of $\delta$ $\sim$ 10, associated with a extremely large and luminous
\lya nebula \citep{Cai2016b}. All these surveys probe redshift slices
of $\Delta z \sim 0.03 - 0.16$, similar to our own $\Delta z = 0.037$
redshift depth.  Although it is difficult to directly compare these
overdensity contrasts with our own values because of different kernels and
field sizes, as well as different $\Delta z$'s, the $\delta_{\Sigma}$ and
$\delta$ of our overdensity are roughly comparable to these proto-cluster
candidates.

\begin{deluxetable*}{lcccccccc}
\centering
\tablecolumns{9} 
\tablecaption{Comparison with other LAE surveys.}
\tablehead{ 
\colhead{Survey}&
\colhead{$z$}&
\colhead{$\Delta z$}&
\colhead{EW cut}&
\colhead{$L$(\lya)}&
\colhead{$C_{\Delta z}$} &
\colhead{$C_{\rm EW}$}   & 
\colhead{$C_{L}$} &
\colhead{$\bar{\Sigma}$} \\
\colhead{(1)}&
\colhead{(2)}&
\colhead{(3)}&
\colhead{(4)}&
\colhead{(5)}&
\colhead{(6)}& 
\colhead{(7)}& 
\colhead{(8)}&
\colhead{(9)}
}
\startdata
Nilsson (2009)  & 2.2 &  0.1061 & 20 & 42.36 & 0.364   & 1.000   & 1.565   & 0.107  \\
Guaita (2010)   & 2.1 &  0.0411 & 20 & 41.80 & 0.940   & 1.000   & 0.415   & 0.075  \\
Mawatari (2012) & 2.4 &  0.0683 & 25 & 41.99 & 0.566   & 1.121   & 0.611   & 0.107  \\
Prescott (2008) & 2.7 &  0.1653 & 40 & 42.18 & 0.234   & 1.580   & 0.955   & 0.350  \\
Palunas (2004)  & 2.3 &  0.0444 & 36 & 42.78 & 0.870   & 1.475   & 8.958   & 0.201  \\
\hline\\[-1.5ex]
This work       & 2.3 &  0.0370 & 20 & 42.19 & \nodata & \nodata & \nodata & 0.074
\enddata
\tablecomments{
(1) reference for the survey,
(2) survey redshift,
(3) redshift depths from filter widths,
(4)--(5) selection criteria for EWs (\AA) and \Lya luminosity $(\log[L/\unit[]{erg\,s^{-1}}])$,
(6)--(8) correction factors for the redshift depth, EW, and \lya luminosity, introduced in Section \ref{comparison},
(9) average surface density (arcmin$^{-2}\Delta{z}^{-1}$) over the entire field corrected for our sample selection criteria.
}
\label{tab:comparedens}
\end{deluxetable*}

\subsection{Will \boo J1430+3522 Evolve Into a Cluster Today?}
\label{sec:cluster_evolution}

Using \lya emitters as a density tracer, we discover an overdense region
with a projected surface density of $\delta_{\Sigma}$ = \sodenspeakwerrors
and a radius of $\sim$10\,Mpc. To address whether this structure could
collapse into a virialized galaxy cluster by $z=0$, i.e., whether it is
in fact a ``proto-cluster'', we compare our observations with the analysis
of structure formation from cosmological simulations by \cite{chiang13}.

Using the Millennium Run [MR; \cite{springel2005}] cosmological
simulation, \cite{chiang13} identify the mass, extent, and density
contrast that galaxy cluster progenitors must have in order to evolve
into galaxy clusters at $z = 0$. In their study, a cluster is defined
as a virialized dark matter halo with a total mass greater than
$10^{14}{M_\odot}$ at redshift $z=0$. Based on this definition they
track the evolution of DM haloes and galaxies in $\sim$3000 clusters
from early epochs ($z$ = 7) to present day.
For this sample, they calculate the correlation between galaxy density
contrast of protoclusters at different redshifts and the mass of
its present-day cluster offspring. They also show how the projected
density contrast is affected by the redshift uncertainty $\Delta z$
of a survey, demonstrating that potential protocluster over-densities
become observationally indistinguishable from the field, for all except
the most massive structures, if the over-densities are measured with
$\Delta z > 0.1$. Thus, wide-field narrowband imaging surveys are the
key to identifying early stages of cluster formation.

In their analysis, \cite{chiang13} measure the density contrast of the
structures in the MR data after smoothing it with $(\unit[15]{Mpc})^3$ and
$(\unit[25]{Mpc})^3$ tophat cubic kernels.  To match these kernel sizes,
we smooth our survey map with $(15\times15\times\Delta z)$\,Mpc$^3$ and
$(25\times25\times\Delta z)$\,Mpc$^3$ rectangular windows, with a redshift
uncertainty of $\Delta z=0.037$ (46.6\,Mpc).  In this configuration,
we find $\delta_{\Sigma,15}$ = 3.4 and $\delta_{\Sigma,25}$ = 2.0 for
\boo J1430+3522.

\begin{figure*}[t]
\centering
\includegraphics[width=0.85\textwidth]{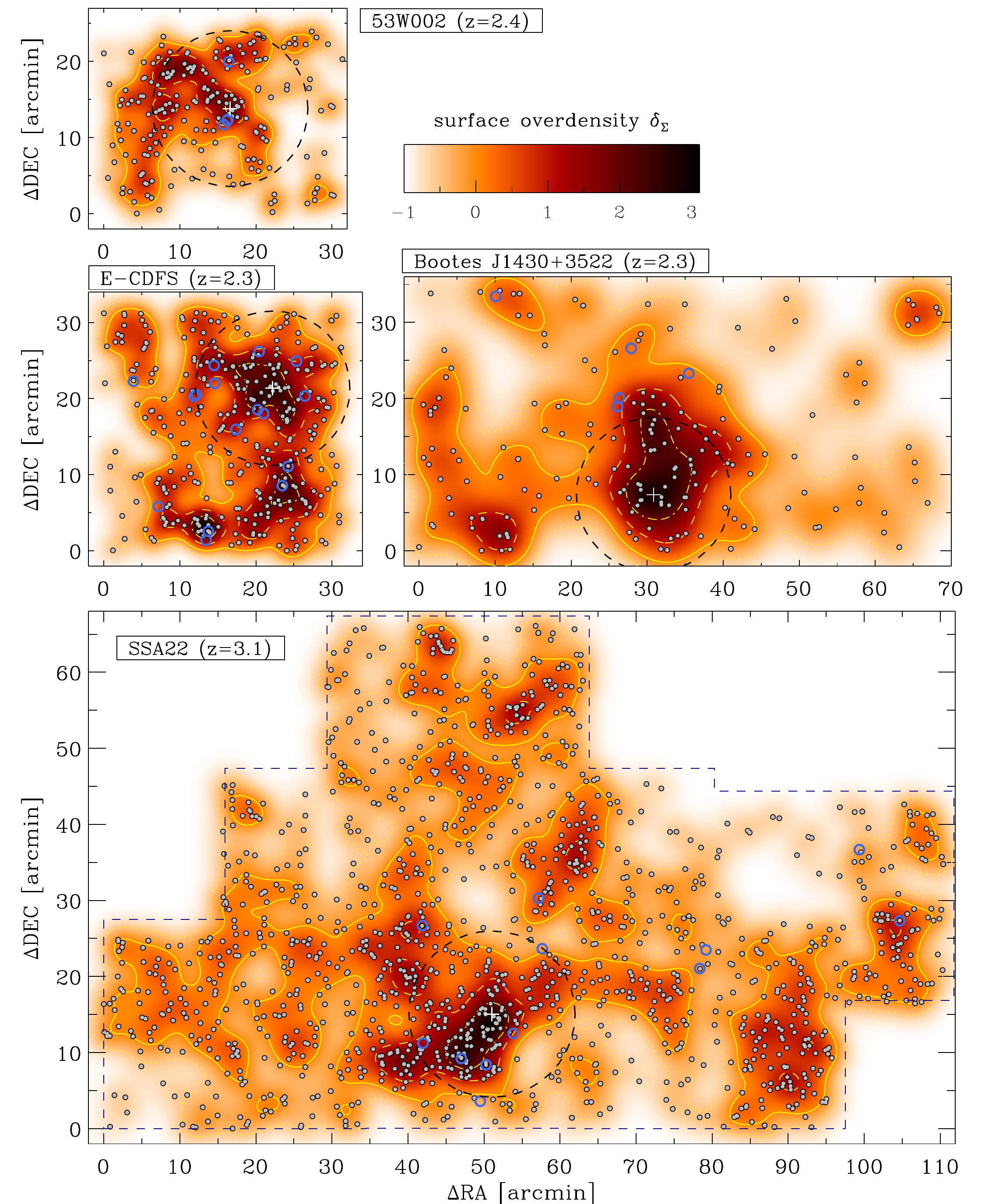}
\caption{
Surface overdensity maps for four survey fields: 53W002
\citep{Mawatari12}, E-CDFS \citep{YY2010}, \boo J1430+3522 (this work),
and SSA22 \citep{M2011,yamada2012}.  \lya emitters and blobs are marked
with gray dots and blue open circles, respectively.  Four contours
(dot, solid, dashed, dot-dashed) represent $\delta_{\Sigma}$
= $(\Sigma-\bar{\Sigma})/\bar{\Sigma}$ = $-$0.5, 0, 1, and 2,
respectively. The dashed black circles are centered on highest peaks of
each region with a radius of 5 physical Mpc. The \Lya blobs often live
on the outskirts of the highest peaks.
The sizes and the peak amplitudes of the overdensities are consistent from
field to field; three overdensities in E-CDFS, \boo J1430+3522, and SSA22
fields have 8.5 -- 10 physical Mpc diameters for the $\delta_{\Sigma}$
= 1 contour (dashed) and the peak $\delta_{\Sigma}$ = 2.8 -- 3.0. Note
that the sizes of the overdensities do not increase with survey size,
which is consistent with them being the largest overdensities at this
epoch and evolving into the richest clusters today.
}
\label{fig:map_kde}
\end{figure*}

According to \cite{chiang13}, an uncertainty of $\Delta z \approx 0.037$
(46.6\,Mpc) in the redshift of the \Lya emitters used to trace an
overdensity at redshift $z$ = 2--3 reduces the apparent surface density
contrast by $\sim$50\% compared to its original value calculated using
$(\unit[15]{Mpc})^3$ cubic windows. This is because with increasing
redshift uncertainties, more galaxies in the background and the foreground
of the overdense regions are included in the analysis and smooth out
irregularities in surface density.  Correcting for this effect, we
obtain a $\delta_{15,{\rm corrected}}$ $\approx$ 6.8.  Note that if we
assume that the overdensity is confined only within the (15\,Mpc)$^3$
cube, $\delta_{15,{\rm corrected}} = 10.6$ would be required to yield
the observed $\delta_{\Sigma,15}$ = 3.4. Therefore, $\delta_{15,{\rm
corrected}}\approx 6.8$ should be a reasonable value for the density
contrast over the (15\,Mpc)$^3$ cubic window.

This density contrast is much higher than $\delta_{15} = 2.88$, the value
needed for a $z\sim2$ structure to evolve into a cluster at $z$ = 0 with
$>$80\% probability \citep{chiang13}.  Here we have adopted $\delta_{15}$
using galaxies with SFR $>$ 1\,$M_\sun$ yr$^{-1}$, which are analogs
to \lya emitter populations.  This $\delta_{15,{\rm corrected}}$ is
high enough for it to evolve into a present-day cluster with near 100\%
certainty, even if a wide range of other tracer populations are assumed
(see Fig.\ 8 of \citealt{chiang13}).  Therefore, we conclude that \boo
J1430+3522 is indeed a ``proto-cluster''.

Finally, using the correlation found by \cite{chiang13} between galaxy
contrast at a given epoch and present day cluster mass, we estimate the
future mass  protocluster to be $\log{(M/M_\odot)}$ $\sim$ $15.1\pm0.2$
similar to that of Coma cluster.

\subsection{Size and Amplitude of Protoclusters}

\begin{figure}[t]
\centering
\includegraphics[width=0.47\textwidth]{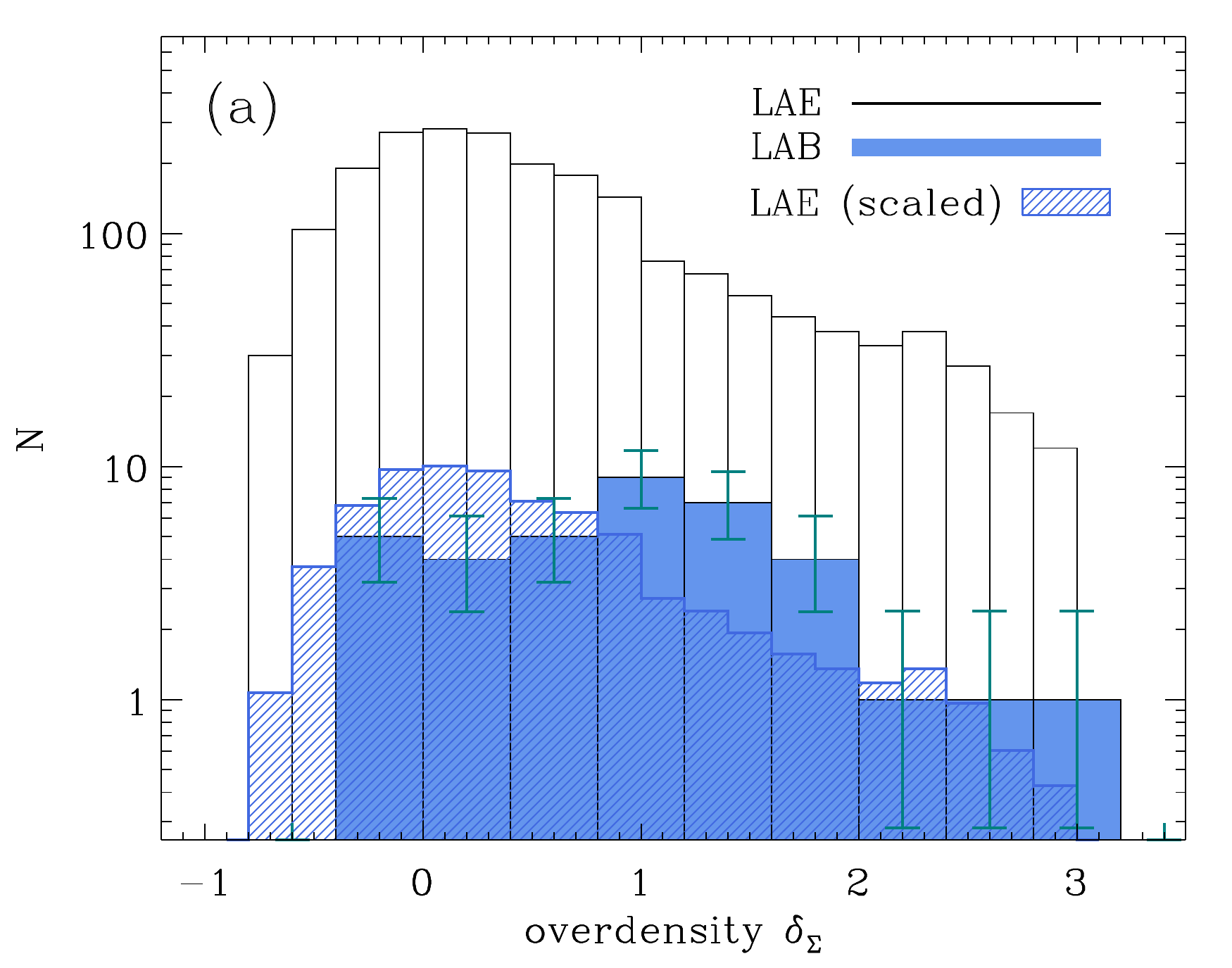}
\caption{
Distribution of surface overdensity $\delta_{\Sigma}$ for \lya emitters
(open histogram) and \Lya blobs (blue filled histogram). The shaded
histogram is the open histogram scaled down for easier comparison with
the \Lya blob distribution.  The \lya blob distribution shows a relative
excess at $\delta_{\Sigma} = 1-2$, suggesting that they favor moderately
overdense regions of LAEs.
}
\label{fig:overdensity}
\end{figure}

We have discovered a new protocluster traced by LAEs and \Lya blobs.
To compare it to other known protoclusters, we compile previous narrowband
imaging surveys at $z$ = 2--3 that have discovered both \lya blobs
and protoclusters in the same field. These three protoclusters are
located in the E-CDFS \citep{YY2010}, the 53W002 \citep{Mawatari12},
and SSA22 fields \citep{yamada2012, M2011} at $z$ = 2.3, 2.4 and 3.1,
respectively.\footnote{\cite{P2008} and \cite{Erb2011} also found
\lya\ blobs associated with overdensities traced by LAEs. However,
the coordinates of the LAEs in their fields are not available, and it
is unknown if the small survey area (220arcmin$^2$) of \cite{Erb2011}
includes the whole overdensity.}

We reproduce the surface density maps of \lya emitters and blobs for
these three fields and \boo J1430+3522 field in Figure \ref{fig:map_kde}.
To make these maps, we use the KDE and cross-validation method presented
in this paper with Gaussian kernel widths of $\sigma$ = 1\farcm43,
1\farcm55, 1\farcm87, and 2\farcm64 for the 53W002, E-CDFS, SSA22, and
\boo J1430+3522 fields, respectively. These kernel sizes are within 25\%
of the values originally adopted by each survey: $\sigma$ = 1\farcm5 for
53W002 \citep{Mawatari12}, 1\farcm2 -- 2\farcm2 for E-CDFS \citep{YY2010},
and 1\farcm5 for SSA22 \citep{yamada2012}. To test the effect of kernel
sizes on our results below, we also produce maps with (1) the values
adopted in each reference and (2) a same width (1.5\arcmin) for all
four fields.  Our results here do not change with the choice of the
kernel size.

Figure \ref{fig:map_kde} shows the contours of surface over-density
$\delta_{\Sigma}$ = $(\Sigma-\bar{\Sigma})/\bar{\Sigma}$ for each survey.
When calculating $\delta_{\Sigma}$, we estimate $\bar{\Sigma}$ over
each survey.  For the E-CDFS protocluster \citep{YY2010,Balestra2000}
which almost fills the 30\arcmin$\times$30\arcmin\ field, we use the
$\bar{\Sigma}$ from our \boo survey because both surveys used the same
narrowband filter and sample selection methods.

Figure \ref{fig:map_kde} shows that both the peak amplitudes and the
sizes of the protoclusters are consistent with each other, despite
the wide ranges of survey areas probed in each survey.  In particular,
three protoclusters in the E-CDFS, \boo J1430+3522, and SSA22 fields have
almost identical peak surface density contrasts of $\delta_{\Sigma}$ = 2.8
-- 3.0.  In contrast, the 53W002 protocluster has smaller size and lower
peak amplitude than the others, suggesting it is only moderately rich.
The three protoclusters (E-CDFS, \boo J1430+3522 and SSA22) have 8.5 --
10 physical Mpc diameters (28--39 comoving Mpc; 17\arcmin--21\arcmin)
if we measure largest dimension of the $\delta_{\Sigma}$ = 1 contour
(dashed).  The linear size of the protocluster does not grow bigger
than this typical size even though the survey area increases from E-CDFS
(35\arcmin), \boo J1430+3522 (70\arcmin) to the SSA22 field (110\arcmin).
For $\delta_{\Sigma}$ = 2 (dot-dashed) contour, the protoclusters also
have similar sizes of 4.6 -- 7.2 physical Mpc (16--24 comoving Mpc)
with wider ranges.

Protocluster overdensity profiles from simulations \citep{chiang13}
show that even for the most massive protoclusters at redshift $z$ = 2--3
(i.e., progenitors of galaxy clusters with a present-day mass greater than
$10^{15}$M$_\odot$), the average diameter of areas with a volume density
contrast above $\delta$ = 1 and 2 is $\approx$32 and $\approx$24\,Mpc,
respectively. Although it is not straightforward to relate the size
measured for a fixed surface density contrast ($\delta_{\Sigma}$) to
that measured for a volume density contrast ($\delta$), these sizes are
in good agreement with the observations discussed above.

The comparable extents of protoclusters at $z$ = 2--3, and the fact
that their observed size does not grow with the extent of the surveyed
field, suggests that they are the largest bound structures at that epoch.
It is clear from our results that a very wide-field survey over $\sim$1
degree is required to reliably confirm massive protoclusters at this
epoch and to determine their full physical sizes.

\subsection{Ly$\alpha$ Blobs in Protocluster Outskirts}

Visually, all the maps in Figure \ref{fig:map_kde} are striking;
the \Lya blobs often lie outside the densest concentration of LAEs.
\citet{Mawatari12} also note that all four of their LABs are
located on the edges of high-density regions. To quantify relative,
local environments of \lya emitters and blobs, we measure their
local over-densities from the smoothed surface density maps (Figure
\ref{fig:overdensity}). The distribution of LAEs' local overdensities is
similar to the lognormal distribution that is known to well approximate
the dark matter distribution \cite[e.g.,][]{Coles&Jones1991,Orsi2008}.  The
two-sample Kolmogorov-Smirnov (K-S) test shows that the distributions of
the LAE and LAB populations are different at the 3.8$\sigma$ significance
level.  The distributions differ most at moderate over-densities,
$\delta_\Sigma = 1-2$, where there is a clear excess of \lya blobs.
While it is not re-analyzed here, the LABd05 blob is also located near
the region of $\delta_\Sigma \sim 1.3$ \cite[][see their Fig.\,3]{P2008}.
Likewise, the six \lya blobs in \cite{Erb2011} appear to lie at the
edges of the HS 1700+643 protocluster field.
We conclude that \lya blobs prefer moderately over-dense regions of
LAEs that are twice or three times denser than the average density of
the survey ($\delta_\Sigma \approx 0$), perhaps avoiding the densest
regions within a protocluster.

Why do \lya blobs occupy the moderate over-dense region or outskirts of
protoclusters? One possibility is that \lya blobs represent proto-groups
that are accreting into a more massive protocluster from the cluster
outskirts.
\citet{P2012} found that the LABd05 \lya blob \citep{D2005} contains
numerous compact, small, low-luminosity ($<$0.1$L_{*}$) galaxies.
Similarly, \citet{YY2011, YY2014b} identify several H$\alpha$
or [\ion{O}{3}] emitting sources within \lya blobs with relative
line-of-sight velocity differences of $\sim$200 -- 400 km s$^{-1}$, which
are consistent with the velocity dispersions of $\sim10^{13}$M$_\odot$
galaxy groups. Furthermore, the number and variance of \Lya blobs is
consistent with them occupying $\sim10^{13}$M$_\odot$ halos.
We speculate that the extended \lya-emitting gas may be the
proto-intragroup medium and/or stripped gas originating from
galaxy-galaxy interactions within these proto-groups.

We test the plausibility of this scenario by checking if the
expected number of proto-groups in the massive protocluster
environment is roughly consistent with that of the \lya blobs
around \boo J1430+3522.  We estimate that our LAE over-density  will
evolve into a $\sim$ $10^{15}$M$_\odot$ rich cluster today (Section
\ref{sec:cluster_evolution}). In this case, simulations predict that
the current protocluster mass is $\sim$ $10^{14}$M$_\odot$ and that
it accretes $\sim$15 $10^{13}$M$_\odot$ halos from $z$ $\sim$ 2.3 to 0
\citep{Gao2004, Giocoli2008, Jiang2016}. Thus, the five \Lya blobs that
we detect within $\sim$10$\,$Mpc ($\sim$5 virial radii) could plausibly
trace some of the group-like halos that build the cluster.

\section{Conclusions}
\label{sec:conlusions}

We carry out a deep narrowband imaging survey of a $\sim$1\degr
$\times$ 0.5\degr\ region at $z$ = 2.3 around a known bright \Lya blob
pair discovered by a blind narrowband survey \citep{YY2009}.  We test
whether bright \lya blobs are indeed a tracer of over-dense regions at
high redshift.

We find a total of \nlaes \Lya emitters including three new
intermediate \lya blobs in our 69.3\arcmin$\times$35.4\arcmin\
field.  The average \lya emitter surface density in our field is
$\bar{\Sigma}$ = \rhosv corresponding to a volume density $n$ = \rhovv
over the survey volume of \volume.  The surface density varies from
5.4$\times$10$^{-2\,}$arcmin$^{-2}$$\Delta z^{-1}$ in the field region
to 0.27 arcmin$^{-2\,}$$\Delta z^{-1}$ at the densest part,  in good
agreement with results from previous surveys that targeted either field
or protoclusters at similar redshifts.

We discover a massive over-density (\boo J1430+3522) of \lya emitters with
a surface density contrast of $\delta_{\Sigma}$ = \sodenspeakwerrors,
a volume density contrast of $\delta$ $\sim$ \vodens, and a projected
diameter of $\approx$\,20 comoving Mpc.  By comparing our measurements
with an analysis of the MR cosmological simulation \citep{chiang13},
we conclude that this large-scale structure is indeed a protocluster
and is likely to evolve into a present-day Coma-like galaxy cluster with
$\log{(M/M_\odot)}$ $\sim$ $15.1\pm0.2$.

In our survey and three others we re-analyze here, the physical extent and
peak amplitude of the LAE overdensities are consistent across the surveys.
Because these properties do not increase with survey size, it is likely
these overdensities are the largest structures at this epoch and will
indeed evolve into rich clusters today.

The discovery of a proto-cluster in the vicinity of the two \lya
blobs, along with the discovery of three new nearby LABs, confirms that
bright \lya blobs are associated with overdense regions of LAEs.  Yet,
among the four surveys we analyze, LABs tend to avoid the innermost,
densest regions of LAEs and are preferentially located in the outskirts
at density contrasts of $\delta_\Sigma$ = 1--2. This result and the
likelihood that blobs themselves occupy $\sim10^{13}$M$_\odot$ individual
halos \citep{YY2010} suggest that \lya blobs represent proto-groups that
will be accreted by the protocluster traced by LAEs. In that case, the
extended \lya-emitting blob gas may be a precursor of the intra-group
medium, and ultimately a contributor to the intra-cluster medium.

\acknowledgements
We thank the anonymous referee for her or his thorough reading of the
manuscript and helpful comments.
We thank Lucia Guaita for providing the \lya emitter catalogues, and
Ashoordin Ashoormaran for his insights on the sub-halo mass function.
T.B.\ and Y.Y.\ acknowledge support from the BMBF/DLR grant Nr.\ 50 OR 1306.
Y.Y.'s research was supported by Basic Science Research Program through
the National Research Foundation of Korea (NRF) funded by the Ministry
of Science, ICT \& Future Planning (NRF-2016R1C1B2007782).
Support for B.M.\ was provided by the DFG priority program 1573 ``The
physics of the interstellar medium''. T.B., A.K.\ and F.B.\ acknowledge
support by the Collaborative Research Council 956, sub-project A1,
funded by the Deutsche Forschungsgemeinschaft (DFG).
A.I.Z. acknowledges support from NSF grant AST-0908280 and NASA grant
NNX10AD47G.

\appendix
\section{Estimating LAE surface density using KDE and Cross-validation}

To build a continuous \lya emitter density map (Figure
\ref{fig:contour}$b$) from the spatial distribution of \lya emitters
(Figure \ref{fig:radecmap}$a$), we use the kernel density estimation
(KDE) method \citep{rosenblatt1956, parzen1962}  with a cross-validation
technique. Assuming that the sky positions of our \lya emitter sample
\{$\vect{x}_1, \vect{x}_2,..., \vect{x}_{\rm N}$\} are randomly drawn
from an underlying unknown surface density distribution $f(\vect{x})$,
our goal is to find an estimator $\hat{f}(\vect{x})$ for the true
distribution. Using KDE 
\begin{equation}
\hat{f}(\vect{x})=\sum\limits_{j=1}^{N}K(\vect{x}-\vect{x}_j;\,\bm{\sigma}_j), 
\end{equation}
where $K(\vect{x};\,\bm{\sigma})$ is a normalized kernel, e.g., in a
functional form of uniform, triangular, or Gaussian.  The \bm{$\sigma$}
is a bandwidth, a free smoothing parameter that strongly influences
the estimate obtained from KDE. Note that \bm{$\sigma$} can be one
or two dimensional, as well as different for each datum. In our
application, we consider 1--D and 2--D Gaussian kernels:
\begin{eqnarray}
K(\vect{x};\,\bm{\sigma}) &=& \frac{1}{\sqrt{2\pi}\sigma}\exp{\left[-\frac{x^2}{2\sigma^2}\right]} \\
K(\vect{x};\,\bm{\sigma}) &=& \frac{1}{2\pi\sigma_x \sigma_y}\exp{\left[-\frac{x^2}{2\sigma_x^2}+\frac{y^2}{2\sigma_y^2}\right]}
\label{eqn:kernel2}
\end{eqnarray}

Our goal is to determine the $\sigma$ that best describes the data
itself. KDE is mathematically identical to smoothing a map image with a
Gaussian kernel, the approach most often taken in the literature, although
the smoothing widths are often chosen rather arbitrarily. We show below
that an optimal $\sigma$ can be determined from the data themselves.
For that purpose, we use a leave-one-out cross-validation scheme
\cite[e.g.,][]{Hogg2008}: Let $\hat{f}_{-i}(\vect{x})$ be the
kernel density estimate of $f$ that is obtained from our sample
excluding the $i$-th element.  The probability of finding that $i$-th
element at the observed position $\vect{x}_i$ is proportional to
$\hat{f}_{-i}(\vect{x}_i)$:
\begin{equation}
  \hat{f}_{-i}(\vect{x}_i) 
= \sum\limits_{j=1;j\neq i}^{N} 
  \frac{1}{\sqrt{2\pi}\sigma} \exp{\left[-\frac{||\vect{x}_i-\vect{x}_j||^2}{2\sigma^2}\right]}.
\end{equation}
We then find the parameters that best {\it predict} the observed data by
maximizing the likelihood of find all \{${\vect{x}_i}$\} for a given $\sigma$:
\begin{equation}
	L(\{\vect{x}_i\}_{i=1}^N|\sigma) 
	= \prod\limits_{i=1}^N \hat{f}_{-i}(\vect{x}_i).
\end{equation}

\begin{figure}
\centering
\includegraphics[height=0.406\textwidth, trim=0cm 0cm 0.5cm 0cm, clip=true]{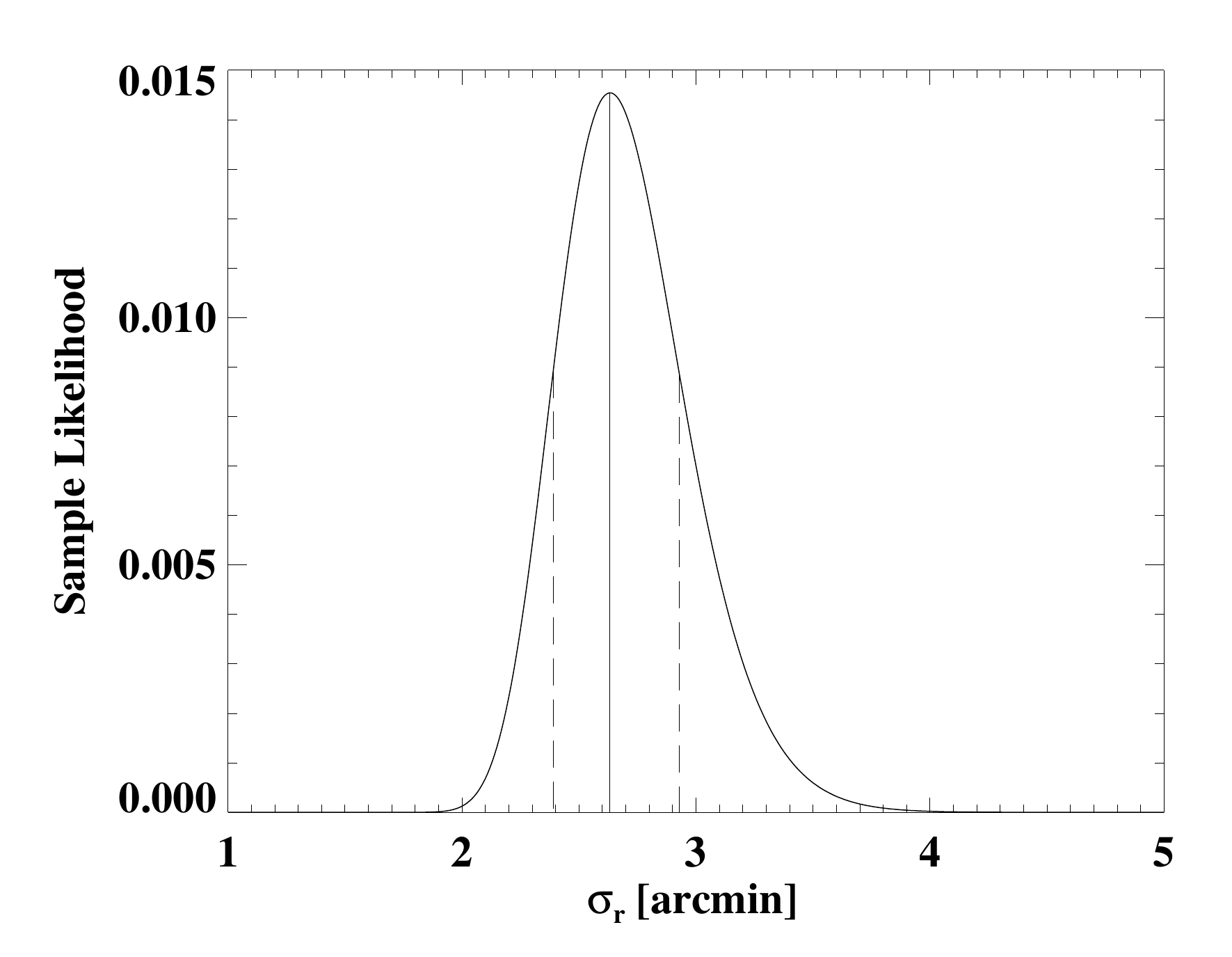} 
\includegraphics[height=0.445\textwidth, trim=1cm 0cm 0cm 1.1cm, clip=true]{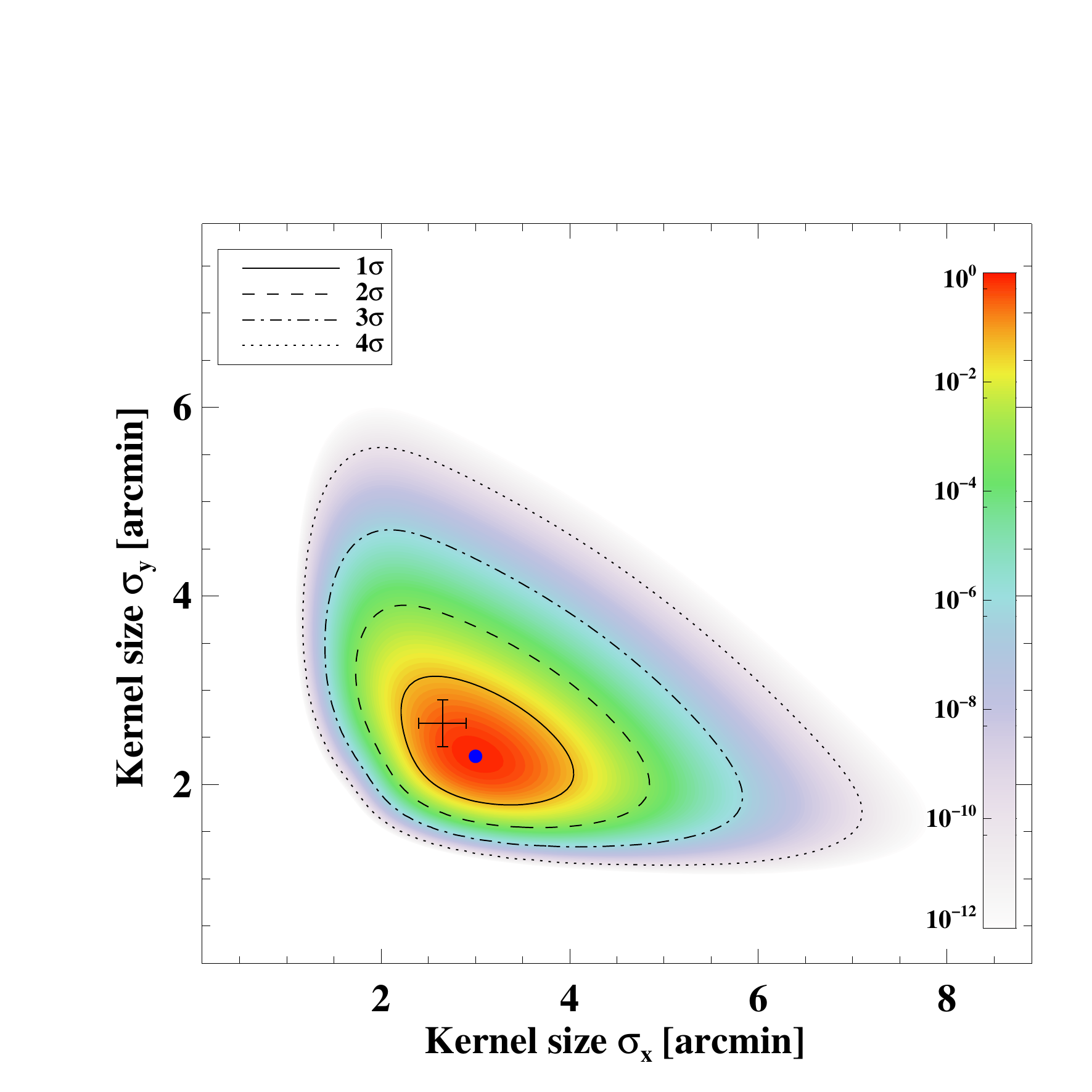} 
\caption{
{\bf (Left)} The sample likelihood function against the width of the
circular Gaussian smoothing kernel $\sigma$.  The vertical line marks
the position of the maximum likelihood at $\sigma$ = 2.63\arcmin\ while
the dashed lines indicate the 1-$\sigma$ uncertainties in kernel size.
{\bf (Right)} Values of the sample likelihood as a function of 2--D
smoothing kernel sizes. The axes show the values of the 2--D Gaussian
kernel widths, $\sigma_{x}$ and $\sigma_{y}$, used for kernel density
estimation. Color levels indicate the values of the likelihood, where
the dot represents its maximum value, while the data point with error
bars indicates the size of a circular Gaussian kernel with $\sigma$
= 2.63\arcmin.
}
\label{fig:kernels}
\end{figure}

We use a simple grid search to determine the kernel width $\sigma$.
Figure \ref{fig:kernels} shows the likelihood $L$ as a function of
$\sigma$ for an 1\arcmin--5\arcmin\ range. The maximum likelihood is
obtained for $\sigma=2.63^{+0.30}_{-0.24}$\arcmin.  If we adopt a 2--D
Gaussian kernel with two smoothing parameters ($\sigma_{x}$, $\sigma_{y}$)
as in Eq.~(\ref{eqn:kernel2}), $\sigma_{x}=3.00^{+1.03}_{-0.78}$\arcmin\
and $\sigma_{y}=2.31^{+0.83}_{-0.52}$\arcmin.  The 1--D kernel width
is within the 68.3\% confidence interval of 2--D kernel width.  We use
$\sigma=2.63$\arcmin\ throughout the paper to estimate the underlying
density distribution.

\bigskip
\bigskip

\bibliographystyle{aasjournal}
\bibliography{./badescu}

\end{document}